%% file: main.tex
\documentclass[]{spie}  

\usepackage{amsmath,amsfonts,amssymb}
\usepackage{graphicx}
\usepackage{glossaries}
\usepackage{sidecap}
\usepackage[colorlinks=true, allcolors=blue]{hyperref}
\sidecaptionvpos{figure}{t}
\title{Approaches to lowering the cost of large space telescopes}

\author[a]{Ewan S Douglas}
\author[b]{Greg Aldering}
\author[c]{Greg W. Allan}
\author[a]{Ramya Anche}
\author[a]{Roger Angel} 
\author[a]{Cameron C. Ard}
\author[d]{Supriya Chakrabarti}
\author[a]{Laird M. Close}
\author[a]{Kevin Derby}
\author[e]{Jerry Edelstein}
\author[a]{John Ford}
\author[f]{Jessica Gersh-Range}
\author[a]{Sebastiaan Y. Haffert}
\author[a]{Patrick J. Ingraham}

\author[a]{Hyukmo Kang}
\author[a]{Douglas M. Kelly}
\author[a]{Daewook Kim} 

\author[a]{Michael Lesser}
\author[a]{Jarron M. Leisenring}
\author[a]{Yu-Chia Lin}

\author[a]{Jared R. Males}
\author[a]{Buddy Martin}
\author[a]{Bianca Alondra Payan}
\author[a]{Sai Krishanth P.M.}
\author[g]{David Rubin}
\author[h]{Sanford Selznick}
\author[a]{Kyle Van Gorkom}

\author[a]{Buell T. Jannuzi}
\author[b,e,i]{Saul Perlmutter}
\affil[a]{University of Arizona (UA), Tucson, AZ, USA}
\affil[b]{Physics Division, Lawrence Berkeley National Lab, Berkeley, CA, USA}
\affil[c]{Massachusetts Institute of Technology, Cambridge, MA, USA}
\affil[d]{University of Massachusetts Lowell, Lowell, MA, USA}
\affil[e]{Space Sciences Laboratory, University of California, Berkeley, CA, USA}
\affil[f]{Princeton University, Princeton, NJ, USA}
\affil[g]{Department of Physics and Astronomy, University of Hawai`i at M{\=a}noa, Honolulu, Hawai`i, USA}
\affil[h]{Ascending Node Technologies, LLC, Tucson, AZ, USA}
\affil[i]{Department of Physics, University of California Berkeley, Berkeley, CA, USA}

\authorinfo{Further author information: (Send correspondence to E.S.D.)\\E.S.D.: E-mail: douglase@arizona.edu}

\begin{document} 
\maketitle

\begin{abstract}
New development approaches, including launch vehicles and advances in sensors, computing, and software, have lowered the cost of entry into space, and have enabled a revolution in low-cost, high-risk Small Satellite (SmallSat) missions. To bring about a similar transformation in larger space telescopes, it is necessary to reconsider the full paradigm of space observatories. Here we will review the history of space telescope development and cost drivers, and describe an example conceptual design for a low cost 6.5 m optical telescope to enable new science when operated in space at room temperature. It uses a monolithic primary mirror of borosilicate glass, drawing on lessons and tools from decades of experience with ground-based observatories and instruments, as well as flagship space missions. It takes advantage, as do large launch vehicles, of increased computing power and space-worthy commercial electronics in low-cost active predictive control systems to maintain stability. We will describe an approach that incorporates science and trade study results that address driving requirements such as integration and testing costs, reliability, spacecraft jitter, and wavefront stability in this new risk-tolerant ``LargeSat'' context. 
\end{abstract}
\input{acronyms}
\newacronym{ESC}{ESC}{ExtraSolar Camera}
\newacronym{SN}{SN}{supernova}
\newacronym{KPNO}{KPNO}{Kitt Peak National Observatory}
\newacronym{NGST}{NGST}{Next-Generation Space Telescope}
\newacronym{IUE}{IUE}{International Ultraviolet Explorer}
\newacronym{TMA}{TMA}{anastigmatic three-mirror}
\newacronym{RFCML}{RFCML}{Richard F. Caris Mirror Lab}
\newacronym{ULE}{ULE}{ultra-low expansion}
\newacronym{OAO}{OAO}{Orbiting Astronomical Observatories}


\section{INTRODUCTION}

\label{sec:intro}  
Spectacular astrophysical results have been produced by large space missions, such as the \gls{HST} and the \gls{jwst}, that grew out of small pathfinder missions\cite{roman_lst_1974,linsky_uv_2018,rieke_last_2021}.
 At present,  development risk in astrophysics is largely concentrated on small, low-cost suborbital or Small Satellite (SmallSat) programs, and ``a too-big-to-fail mentality pervades agency thinking when it comes to NASA's larger and most important missions."\cite[Mr. Paul Martin, Inspector General, NASA]{smith_nasa_2018}. 

This discrepancy mirrors the launch vehicle landscape before the recent precipitous decline in launch vehicle costs. 
\begin{SCfigure}
    \centering
    \includegraphics[width=0.7\textwidth]{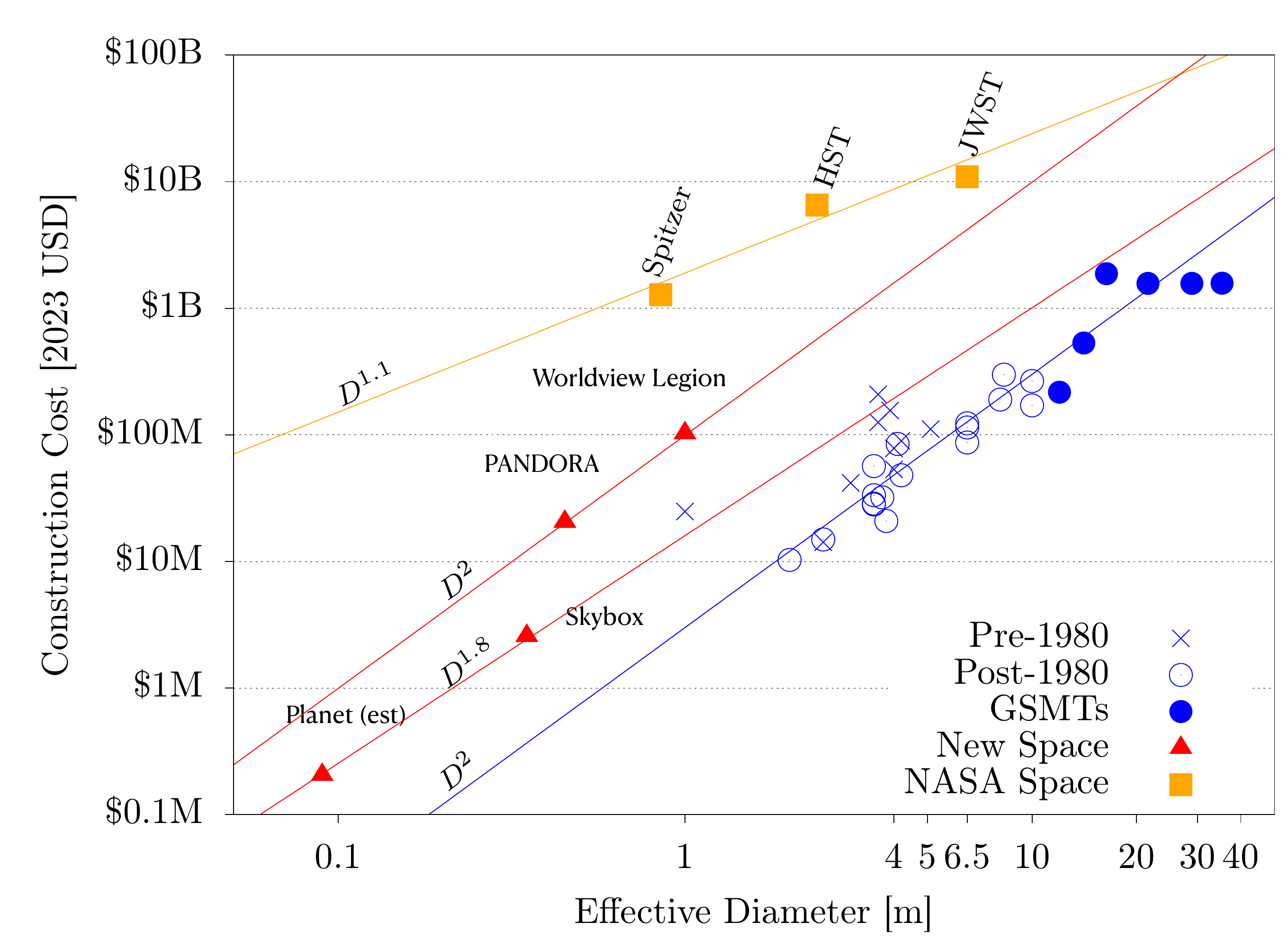}
    \caption[Telescope Costs]{Telescope costs, estimated from public sources, as a function of primary mirror diameter.
    Ground based observatories are fit with  the bottom trend line, blue, and have a significant decrease after 1980 when segmented and honeycomb mirror technologies were introduced (circles). Three classes of space observatories are shown,  the bottom D$^{1.8}$ trend-line shows production line NewSpace SmallSats (for downlooking imaging) while the 3rd line shows limited production NewSpace both for research\cite{quintana_pandora_2021} and imaging.
    All these costs are small compared to the NASA flagship costs (squares), which are likely flatter due to greater baseline costs.  Ground-based telescope costs from Van Belle and Meinel
    \cite{2004SPIE.5489..563V} and online sources\footnotemark, space telescopes costs from various sources\cite{2021Natur.600..208W,jwst_icrp}
    \protect\footnotemark
    \protect\footnotemark .
    \label{fig:telcosts_20230830}
    }
\end{SCfigure}
Elvis, Lawrence, and Seager (2023) \cite{elvis_accelerating_2023} recently described how the combination of decreased launch costs, and increased lift mass and volume enabled by the SpaceX Starship has the potential to transform astrophysics mission design.

While it is unlikely that space-borne observatories will fall below the cost of ground-based observatories, Fig. \ref{fig:telcosts_20230830} shows that telescope research into honeycomb\cite{angel_manufacture_1982} and segmented aperture telescopes\cite{nelson_construction_1988}  shifted  the ground-based cost-curve down by over a factor of two after $\sim$1980, and NewSpace\cite{martin_newspace_2017} economies of scale have recently enabled very low-cost, albeit small, imaging missions.
Thus,  research and new economies of scale enabled by prior research into optics, commercial electronics, and the SpaceX StarShip could have a similar impact on space astronomy and drive the cost of multiple large observatories down into the regime between the NewSpace projections and ground-based observatories.
This study is inspired by pathfinding work over the past several decades on   space telescope design and cost-efficient academic, government, and commercial SmallSat and CubeSat missions.
A small portion of that work is described below; which combined with technological advances (Sec \ref{sec:enablingtech}) and a culture of openness (Sec. \ref{sec:reliability}) is beginning to merge with recent experience in SmallSat development in academia. This merger is enabling University-led large space missions, potentially bringing down cost and increasing dissemination of research by integrating engineering and science teams. 
An examples of such integration is the OSIRIS-REx Camera Suite (OCAMS) -- an instrument which sucessfully imaged the asteroid Bennu after a deep space flight. The OCAMS  team in the UA Lunar and Planetary Laboratory worked with Steward Observatory for systems engineering and mechanical support and the College of Optical Sciences for the optical train, with inputs from Utah State University Space Dynamics Lab and others\cite{rizk_ocams_2018}. Similarly, the Aspera SmallSat mission\cite{chung_aspera_2021} science payload is being built by a university collaboration led the University of Arizona and the spacecraft is being built by the University of Toronto.
 This manuscript will provide a brief introduction to past studies of relevant large space telescope designs, establishing the field as an academic research area, and then describe a mission concept currently being studied by a team from several US universities.

\footnotetext{\url{https://elt.eso.org/about/faq/\#question_9}}
\footnotetext{\url{https://www.jpl.nasa.gov/news/press_kits/spitzer/quick-facts/}}
\footnotetext{\url{https://www.planetary.org/articles/cost-of-the-jwst}}

\subsection{History of Astronomical Space Telescope Research}
Design, simulation, and qualification of optical systems for launch requires a process of research into materials, mechanical, thermal, controls, and structural engineering.

Nancy Grace Roman, Lyman Spitzer, and Aden Meinel were among the early visionaries who saw  in the first decade of the space age the great potential for space observatories. Spitzer\cite{spitzer_space_1960,spitzer_beginnings_1962} established the basic physical parameters of a large space observatory and its astrophysical motivation.
Roman\cite{roman_lst_1974,roman_aip_1980} developed an observatory operational concept around guest observers with near real-time control and a rotating suite of instruments, inspired by the \gls{KPNO}.
There were many learning experiences from early missions, such as the\gls{OAO}\cite{hammond_ultraviolet_1970,shockey_study_1981,roman_aip_1980} launched between 1966-1972 with instruments developed at  University of Wisconsin-Madison and Princeton, \gls{IRAS}\cite{aumann_infrared_1977} launched in 1983, \gls{IUE}\cite{boggess_iue_1978} launched in 1978 and \gls{HST}, advancing technologies and concepts of operation for future flagship astronomical observatories larger than 1 m became an established area of applied research at universities across the US and world.

\
Research into the launch worthiness of large space observatories dates back to the beginning of the space age.
In 1962, Meinel described ``high resolution optical telescopes'' including the \gls{OAO} 0.9m telescope \cite{meinel_high_1962}  and succicently described the challenge of building a simple system that survives launch: ``During the launch into orbit a telescope may be subjected to vibration of approximately 5 to 10 g's over a frequency range of 5 to 1500 cycles per second. As a consequence, either the engineer must find a design
that will preserve optical collimation  or the astronomer will have to realign his optical systems after the telescope arrives in orbit.''\cite{meinel_new_1961}
In the 1980s, at the University of Arizona (UA) the Space Infrared Telescope Facility (SIRTF, now known as Spitzer) team studied different approaches to calculating the stresses on the 1 m telescope mirror and published details of the design, \gls{FEM} approach, and compared methods for calculating the probability of launch survival for different support designs\cite{richard_dynamic_1988}.  
In the early 2000s, also at UA, part of \gls{NGST} technology development \cite{baiocchi_demonstration_2001,baiocchi_design_2004,baiocchi_optimized_2004} 0.5 m and 2.0 m lightweight active mirrors were built and tested including acoustic tests of prototypes.

Detailed \gls{FEM} of the vibro-acoustic  survival of active 1-m class space telescope segments have been studied at MIT \cite{cohan_vibroacoustic_2011}. 
Similarly, the response of JWST mirror segments to launch level acoustics and vibrations was modeled, tested and published \cite{saif_nanometer_2015}. 
The European Space Agency (ESA)'s large aperture telescope technology (LATT) project \cite{briguglio_development_2017} built a 0.4 m active telescope optic for space and qualified it for space to \gls{TRL}-5 via measurement of its response to launch loads. Overall, \gls{FEM} approaches have improved our confidence that optical designs will  survival launch, prior to test.

Control of large space structures, particularly line-of-sight pointing, is another key area of research advanced by US universities to enable large space observatories. 
For example, researchers at Texas A\&M University have built and published a detailed reaction wheel model to improve pointing control of large spacecraft \cite{junkins_near-minimum-time_1991}.
Similarly, an MIT-led series of experiments summarized by Saenz-Otero (2005)\cite{saenz-otero_using_2005} include the Middeck 0-Gravity Dynamics Experiment \cite{crawley_middeck_1993} and Middeck Active Control Experiment (MACE)\cite{miller_middeck_1995}, which tested dynamics and control of large structures required to control the attitude of large space observatories in flight aboard the Space Shuttle. 


Thermal stability of large space telescopes is an essential technology for space astronomy, enabling long-duration exposures on faint objects without changes in the image quality or the distribution of starlight.
Examples include the 1.5-m AMTD-2 mirror, which was studied theoretically \cite{brooks_advanced_2015-1} and then qualified at picometer level stability in a vacuum cryo-shroud simulating space \cite{brooks_precision_2022}.

Material selection is a key aspect of telescope design and many candidate materials are typically studied before the optimal observatory design for a particular astrophysical question to be established.
\gls{jwst} went through an extensive study of options \cite{stahl_jwst_2004} and published details of the cryo-vacuum testing \cite{stahl_jwst_2007}. 
The \gls{jwst} primary mirror segments are Beryllium but several other materials were considered and studied in active research programs, including \gls{ULE} glass at Kodak \cite{matthews_kodak_2003} and Zerodur \cite{baiocchi_demonstration_2001} at the University of Arizona. 
The Kodak AMSD mirror program studying ULE options, including wavefront error in cryogenic vacuum, were published \cite{matthews_kodak_2003}.    
The 3.5 m SiC Hershel telescope was cryogenically tested and discrepancies between the \gls{FEM} and the physical performance of the system were identified, studied, and published \cite{catanzaro_herschel_2009}.

West et al. 2010\cite{west_space_2010} combined some of these development areas and studied the launch survival of 4 m borosilicate primary mirrors and introduced the concepts of thermal figuring for honeycomb borosilicate space mirrors and the use of off-axis stars for guiding and wavefront sensing. This design was recently cited in the conceptual design of a 20 m space telescope \cite{eads_20_2020-1} on the Moon. 
A similar 50-100 m concept is under study at Paris Observatory \cite{schneider_owl-moon_2022} and these concepts prove key to the simpler, 6.5m space telescope concept discussed below. 

Closely related to thermal control and observatory design is wavefront sensing and control techniques to maintain the alignment and prescription of optical surfaces and thereby preserve image quality. 
A range of scientific studies have investigated optimal wavefront sensing and control approaches for large space telescopes, including ultra-stable 6\,m telescopes with active primary mirrors \cite{martin_next-generation_2022}, 9.2 m  space telescopes with formation-flying laser guide stars \cite{marlow_laser-guide-star_2017,douglas_laser_2019,pogorelyuk_laser-guided_2022}, and active space telescopes with predictive thermal modeling \cite{Gersh-Range-JWST_thermal}.
There have been several research missions to test approaches to wavefront control and the TRL of these approaches for future missions. The PICTURE sounding rocket flights in 2011 and 2015  demonstrated piezo-electric fine pointing and power-on of a \gls{MEMS} deformable mirror in space \cite{mendillo_flight_2012,douglas_wavefront_2018} and the Deformable Mirror Demonstration Mission CubeSat\cite{cahoy_wavefront_2013-1,douglas_small_2021} launched in 2020 was a prototype space-\gls{AO} system that performed closed-loop wavefront control using a \gls{MEMS} mirror\cite{morgan_-orbit_2022}. 
The methods for simulation of large optical system performance, with and without wavefront control, are another active research area that has been significantly advanced by \gls{jwst} and the Roman Coronagraph \cite{perrin_simulating_2012,greenbaum_-focus_2016,mennesson_roman_2022,poberezhskiy_roman_2021,CGISim_manual,ref:GershRange_SPC,ref:GershRange_OS11tool} including through testing  \cite{dean_phase_2006,perrin_james_2014} and on-orbit performance \cite{lajoie_year_2023}. 
These parallel efforts to bring precision wavefront sensing and control to both small  and flagship missions have left us with a wide-range set of tools to maintain telescope optical performance in space. 
The next section will detail several industrial developments that can be leveraged to increase the efficiency of space observatory development.

\begin{figure}
    \centering
    \includegraphics[width=0.7\textwidth]{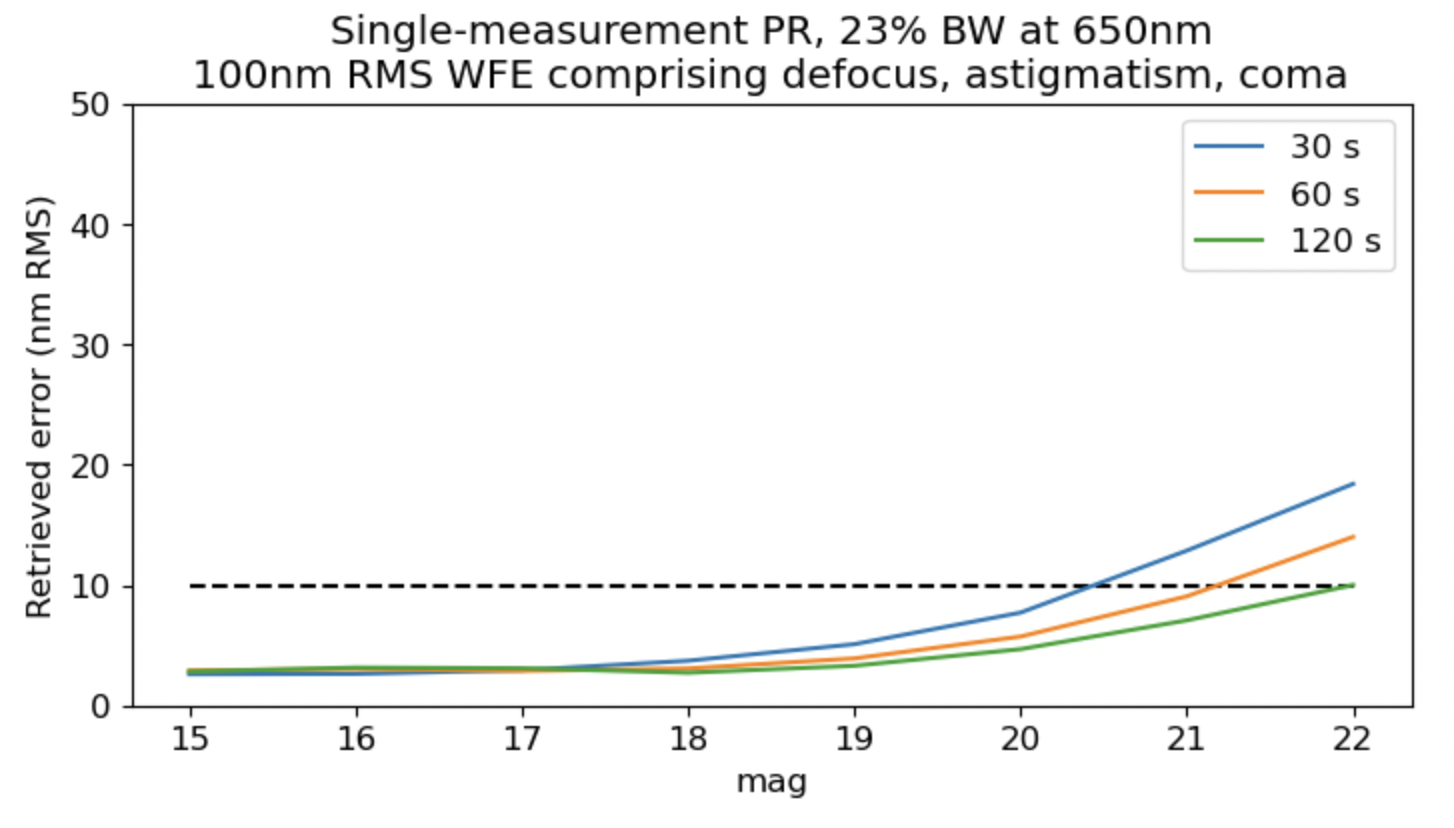}
    \caption{Single-star phase retrieval wavefront eror versus stellar magnitude for a modern CMOS sensor (e.g. IMX571 or IMX455 operating at $\sim$1$^\circ$ Celsius).}
    \label{fig:kyle_phase_1psf}
\end{figure}

\setcounter{footnote}{0}

\section{Enabling Technologies}\label{sec:enablingtech}

\subsection{Launch Cost and Payload Volume}
Spacecraft launch costs and their impact on design are difficult to quantify from public data, but a recent analysis by CSIS\cite{williams_boost-phase_2022} suggests a $>40\times$ decrease in heavy lift launch costs between the Shuttle era and the SpaceX Falcon Heavy; however, the gain decreases to $\sim3\times$ for the ratio between the Saturn V and Falcon Heavy. 

The fairing of the SpaceX Starship has the potential for a 6.5 m JWST-class telescope to be directly launched with a monolithic mirror, removing the cost and complexity of segmented mirror designs. Additionally, while the relative cost of Starship is undetermined,  costs as low as \$10/kg and prices as low as $~\sim$100/kg to LEO have been widely reported, a $>>10\times$ decrease versus Falcon Heavy\cite{scoles_prime_2022,williams_boost-phase_2022,chang_spacex_2023}.

\subsection{Sensor Technology}
The steady advance of silicon sensor technology, driven by economies of scale, with billions of devices manufactured per year \cite{fossum_invention_2020}, has reduced sensor noise and increased pixel count significantly over the past several decades.
For example, the Hubble 1024 $\times$ 1024 \gls{STIS} \gls{ccd}  read noise is 4.4 -- 7.8 e$^-$\,pix$^{-1}$ and dark signal is 0.003\,e$^-$\,sec$^{-1}$\,pix$^{-1}$ at $-86^{\circ}$ Celsius, whereas a modern \gls{COTS}  \gls{CMOS} sensor has 1 -- 3.4\,e$^-$\,pix$^{-1}$ and reaches the same dark signal per pixel at  $0^{\circ}$Celsius with as many as 150 megapixels\cite{alarcon_scientific_2023}. While  CMOS pixel sizes continue to decrease, modern silicon processing advances can be expected to continue to reduce dark current per unit area to improve overall signal-to-noise. CMOS sensors are inherently more radiation tolerant than CCDs and so are well suited for space missions\cite{Goiffon_Vincent}. Modern CMOS sensors also allow faster frame rates than previous generation of silicon imagers. Finally, these devices are being fabricated with increasing physical areas and at higher yields for effectively lower cost.   

\subsection{Radiation Tolerant Computing}
Historically purpose built flight computers ran assembly language and required costly and niche software development skills.
Currently however,  advances in \gls{COTS} computing hardware and software, particularly increases in processing power and radiation tolerant computing, has allowed the deployment of widely supported open-source operating systems such Linux in space\cite{leppinen_current_2017}, decreasing development time while allowing highly complex control systems at relatively low-cost, with recent demonstrations including the operation of the Ingenuity Mars Helicopter\cite{canham_mars_2022} and closed-loop jitter and wavefront control on CubeSats running Linux\cite{knapp_demonstrating_2020,smith_-orbit_2018,morgan_-orbit_2022}.

\subsection{High Earth Orbits}
The Sun-Earth L2 point has become the destination for space observatories requiring a thermally stable environment, with the price being increased communication power requirements and propulsion $\Delta$V. Recently, the TESS mission\cite{ricker_transiting_2014}  demonstrated\cite{parker_transiting_2018} an alternative: a 13.7 day period, high Earth orbit (HEO) in 2:1 resonance with the Moon, reached via lunar gravity assist. The TESS HEO orbit provides a large continuous sky coverage in a thermally stable low radiation environment for relatively low $\Delta V$\cite{gangestad_high_2013}, lowering the propulsion needs for space observatories and increasing potential data downlink relative to L2 orbits for the same transmitter power.

\subsection{Large Optic Fabrication and Design Optimization}
The cost of manufacturing large optical systems depends heavily on the primary optic material. 
Honeycomb borosilicate mirrors cast by the \gls{RFCML} at UA at 1165$^\circ$ C \cite{oswalt_honeycomb_2013} are relatively affordable to manufacture in large apertures, with a complete 6.5\,m ground-based observatory costing of order \$70M USD in 2020 dollars \cite{schlegel_astro2020_2019}.
These mirrors, combined with the \gls{TMA} optical design proposed by Korsch\cite{korsch_anastigmatic_1977}, and recently demonstrated on JWST, enable all-reflective large-field-of-view telescopes\cite{feinberg_space_2012} which can continuously observe a significant number of stars in parallel over a wide range of wavelengths.

\subsection{Phase Retrieval} When Hubble was launched with polished-in spherical aberration, phase retrieval using starlight to determine the optical wavefront error, which would be corrected by repair missions, became a critical area of research\cite{roddier_combined_1993,litvak_image_1991,krist_phase-retrieval_1995}. These families of algorithms were later extended for wavefront sensing of the JWST observatory\cite{dean_phase_2006,lajoie_year_2023,schlawin_jwst_2023}.Phase retrieval allows a wide-field science instrument to operate as the wavefront sensor, extending the concept of using the science instrument for guiding pioneered by TESS and planned for Roman Space Telescope\cite{nguyen_fine-pointing_2018,bartusek_nancy_2022}.

\subsection{Project, Interface, and Document Management}
Over the past several decades, advances in development processes for software with version control, project management, test-based design, and continuous integration and deployment have increased the development pace of increasingly complex software \cite{parnin_top_2017}. With strategic management, integrated software and hardware teams can cycle through the development process faster \cite{tibazarwa_strategic_2021}.  
Breaking down the communication and cultural barriers between software and hardware teams in astronomical space instrument development has the potential to create smaller, more agile teams. As shown by experience gained through the Vera C. Rubin Observatory and the 
Gaia mission, a project can facilitate better communication is by using machine readable interfaces, specifically for software development and simulation teams, but also for disseminating commonly referenced numbers such as the fundamental parameters describing the optical design. E.g.  using the code (or files read in by the code) as the interface control document \cite[ and references therein]{omullane22a} and utilize JSON Schemas\footnote{\url{https://json-schema.org/}} to define the interface structure and definition. This enables the input parameters themselves to be verified upon population using standard software packages (e.g. \href{https://json-schema.org/}{jsonschema in Python}) and offers the ability to populate useful metadata available directly with each parameter, eliminating error-prone manual extraction of variables from definition documents. This structure of ICD is especially useful for performance simulations, particularly when evaluating the impact of potential and/or upcoming changes.

Document management has long been a cornerstone of developing reliable aerospace and astronomical instruments \cite{holm_making_2002}.
One existing trial of such an approach was the \textit{git} based requirements management used early in the Roman Coronagraph science investigation teams \cite{douglas_wfirst_2018}, where systems engineers adopted a software requirements tool (Doorstop\cite{browning_doorstop:_2014}) to improve documentation change tracking and automated distribution.
The natural next step is tying these broad systems requirements to test plans, linking them in a \textit{git} tracked system, and using a code-review-like process to develop both hardware and software tests. These tools make an iterative, prototype-heavy design process more feasible since lessons learned are captured naturally.

\section{Conceptual Design}
We seek to bring  together the I) well established history of space telescope development, II) the recent NewSpace/SmallSat revolution, and III) recent technological advances in hardware and software, into a point design which can be iterated upon, providing a starting point for concept development and improvement leading to the building of prototype(s) to meet the specific requirements of space astronomy, particularly cosmology and exoplanet direct imaging.
The proliferation of CubeSats and SmallSats have shown that in a relaxed design environment with few hard constraints (mass, volume, power), engineering creativity opens.
Decreased launch costs and the successes of CubeSats and SmallSats can overwhelm the mission designer with possibilities. 
To overcome this challenge, we set ourselves the constraints of two challenging toy problems, I) precision spectrophotometry of faint sources such as Type~Ia supernovae, and II) high-contrast imaging of sub-Neptune planets around nearby stars. Further constraints to focus the designers' imagination are: III) use of the publicly available fairing of the SpaceX starship launch vehicle\footnote{\url{https://www.spacex.com/media/starship_users_guide_v1.pdf}} and IV) use the field-proven  Richard F. Caris Mirror Lab 6.5\,m light-weighted borosilicate honeycomb mirror without modification \cite{olbert_casting_1994,martin_fabrication_1997,geyl_polishing_1999,kingsley_inexpensive_2018,miyata_university_2022}. 

The sections below will highlight key aspects of a low-cost, large monolithic observatory conceptual design. Sec. \ref{sec:active} will introduce the \gls{TMA} optical design, which is discussed in greater detail by Kim et al.\cite{kim_compact_2023}, with the alignment error budget presented by Choi et al.\cite{choi_approaches_2023}, while Derby et al.\cite{derby_integrated_2023} describes an integrated modeling of the observatory using continuous a wavefront control approach, which is studied in the lab by Kang et al.\cite{kang_focus_2023} and enabled by continuous control of mirror bending modes, which are described by Blomquist et al.\cite{Blomquist_analysis_2023}.

\subsection{Active Optical Design}\label{sec:active}
To use a borosilicate mirror on the ground requires active control to minimize thermal and gravitationally-induced wavefront gradients\cite{oswalt_honeycomb_2013}.  
Similarly in space, thermal gradients and distortions due to the support structure or gravitational release must be corrected via actuators and/or thermal figuring. 
To design a system that can correct these errors, they must be predicted with sufficient margin prior to launch. 
Even for a monolithic telescope, on-orbit measurement and correction allows relaxation of model accuracy, thereby lowering the costs of model validation and test campaigns, which are significant drivers of large observatory costs.
Lyman Spitzer\cite{spitzer_space_1960,spitzer_beginnings_1962} established the basic physical limitations and derived the natural guide star limitations on guiding the attitude of a space observatory.
This approach was applied by the \gls{HST} \gls{FGS}\cite{nurre_hubble_1989}. 

Following the scaling relations described in Hill et al (2013) \cite[\S 1.5.7]{oswalt_honeycomb_2013} we can approximate the thermal requirements for a borosilicate primary mirror.
A typical UA mirror  figure is $\sim$14 nm RMS surface, 28 nm RMS \gls{WFE}, after removal of low-order modes \cite{geyl_polishing_1999}.
For general astrophysics, to maintain  diffraction-limited imaging (see Choi et al.\cite{choi_approaches_2023}) at 1\,$\mu$m one could allocate  $\sim$30 nm RMS \gls{WFE} to thermal distortions and sensing accuracy of $\simeq$10 nm.
The \gls{CTE} ($\alpha$) of  Ohara E6 borosilicate at room temperature us $\alpha=$ 2.78 $\times10^{-6}/^\circ$C).
Characteristic lengths of the mirror range from the phase sheet (27 mm), the cell center spacing (192\,mm), and the overall thickness $h=$711\,mm \cite{martin_fabrication_1997}.
The last dimension leads to a vertical thermal stability requirement  throughout the mirror given by:
\begin{equation}
\delta T=\frac{\rm{WFE}}{\alpha h}=\frac{10\, nm}{\alpha h}
\end{equation}
Fig. \ref{fig:kyle_phase_1psf} shows for a 6.5\,m telescope with modern \gls{COTS} CMOS sensors, a single 22nd magnitude star allows sensing of low order Zernike polynomials to 10\,nm WFE in 120 seconds. 
Assuming a conservative ground-based \gls{AO} controller bandwidth of 10$\times$ the sampling rate (i.e. 20\,min), thermal drifts of $\sim$0.25\,mK/min (or $\sim$0.5 nm/minute) can be sensed and controlled with a single 22nd\,mag star.

To estimate the \gls{FOV} required to always have two stars sufficiently bright for monitoring low-order aberrations, we take the dim star counts at the poles from Bahcall\cite{bahcall_universe_1980}, reject the binary fraction (~$\sim$50\%) and find that a $>10$ arcm$^2$ wavefront sensor area \gls{FOV} is required.
In order to have margin on sensor failure, Poisson fluctuations in the star counts and enable other science channels to coexist with the incomplete fill-factor of \gls{COTS} sensor packaging, this was translated into a $>50$ arcmin$^2$ focal plane, described in detail in Kim et al\cite{kim_compact_2023}.

\subsection{Fault Tolerant Instruments}
A 6.5\,m telescope in space opens up new scientific paradigms.
The photometric stability of a space telescope allows high-precision photometry much beyond that of ground-based observatories. Even low-cost nanosatellites have made contributions to the literature \cite{elliott_5_2021,knapp_demonstrating_2020} and a 6.5\,m space observatory with even basic instruments has groundbreaking potential. 
In this section, we describe instruments that go beyond the basic imaging required for photometry and continuous wavefront control described in the last section and put forth affordable instrument concepts that leverage the space platform to address unanswered questions in cosmology and exoplanet research.
\subsubsection{Two-octave spectrograph}
A single high-bandwidth, moderate resolution spectrograph with continuous coverage from 0.4--1.7 $\mu$m allows measurement of a wide range of astrophysical phenomena and is particularly useful for the precision history and rate of the expansion of the universe out to redshift $z=2$ and can thereby track the evolution of dark energy through precision spectrophotometry\cite{perlmutter_key_2019}, for instance using the techniques of Boone et al (2021)\cite{boone_2021a, boone_2021b}  without requiring cross-calibration of multiple instrumemts. 
Table \ref{tab:baselineifs} describes a set of instrument requirements to accomplish this science using an Integral Field Spectrograph (IFS) instrument much like those found on ground-based telescopes\cite{peters-limbach_optical_2013,claudi_sphere_2010,lantz_2004, bacon_2010} or previously proposed for space\cite{Prieto_2008, WFIRST_IFS}.

\begin{table}[ht]
    \centering
    \begin{tabular}{p{1.5in}|p{1.5in}|p{3in}}
\hline
\hline \\[-1.1em]
Parameter  &  Value   & Notes  \\[0.3em]
\hline \\[-1.15em]
Wavelength range & $\lambda$= 400\,nm to 1700\,nm& Allows Type\,Ia \gls{SN} characterization at redshifts $0<z<2$\\[0.3em]
Wavelength resolution & R=100 at 1$\mu$m &  S/N $>$ 10 per resolution element (in the rest-frame B band for a flux of AB=25\,mag)\\ [0.3em] 
\gls{FOV} & $\sim$1 arcsecond & Allows spectrophotometry, including subtraction of host galaxy from \gls{SN} spectra\\[0.3em]
Photometric accuracy & 3 millimag & Relative to standard stars \\  
\hline
\end{tabular}
\vspace{0.3em}
    \caption{Desired Spectrograph  Instrument Parameters.}
    \label{tab:baselineifs}
\end{table}

\subsubsection{Extra-solar camera}
\begin{table}[ht]
    \centering
    \begin{tabular}{p{2in}|p{2in}|p{2in}}
\hline
\hline
Parameter  &  Value   & Notes  \\
\hline
     Clear Entrance  Aperture & 2m-2.5 m and 0.6m-1 m & independent  sub-apertures  \\
    Design life  & 1+ year (3+ goal)& sets radiation sensitivity\\
\hline
     Primary operation wavelength & 650 nm &\\
Nominal filter bandwidth & 2\% & minimizes sensitivity to \acrshort{WFE}\\
\hline
Deformable Mirror Actuator Count & 952 & BMC Kilo-C DM 1.5um \\
Deformable Mirror Actuator Stroke (max) & 1500 nm & BMC Kilo-C surface stroke for 4x4 actuators\\
Coronagraph mask & Charge-6 \acrshort{VVC}\\
\acrshort{IWA} & 2.4$\lambda/D$ &depends on mask \\
\acrshort{OWA} & 10$\lambda$/D-15$\lambda$/D & depends on WFE\\
Sensitivity & 1e-8 or dimmer star-planet flux ratio & sensitivity to debris disks and sub-Neptune size planets in habitable zones of nearby stars\\

\hline
\end{tabular}
    \caption{Desired \acrshort{ESC} Coronagraphic Instrument Parameters.}
    \label{tab:baselineesc}
\end{table}

The baseline parameters of a simple coronagraph or \gls{ESC} design (Table \ref{tab:baselineesc}) are designed around unobstructed sub-apertures to simplify coronagraph design and construction, maximizing science per dollar. The straightforward single \gls{DM} instrument layouts are inspired by the CDEEP/SCoOB and PICTURE-C designs\cite{mendillo_picture-c_2019,maier_design_2020,ashcraft_space_2022,van_gorkom_space_2022}. \gls{ESC} concept is built around survey bright stars for reflected light from planets and the the stellar flux enables sensing of the telescope \gls{WFE} much more precisely than in the general astrophysics example above, enabling control down to the sub-nanometer regime required to reach $10^{-8}$ contrasts with a simple charge-6 vector vortex coronagraph and continuous dark hole wavefront sensing, a concept described in more detail in Derby et al\cite{derby_integrated_2023}.

\subsection{Parameterized disturbances}
\begin{figure}[htbp]
\begin{center}
\centering
\includegraphics[width=0.7\textwidth]{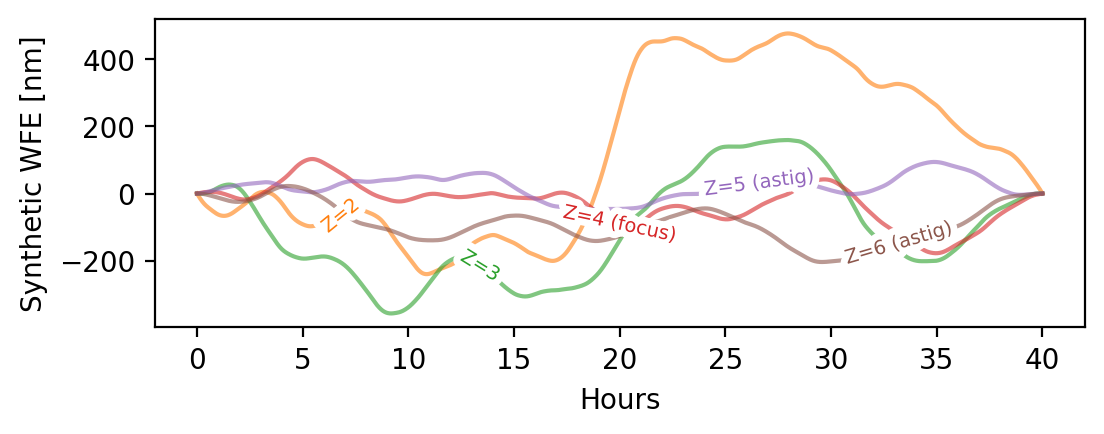}
\includegraphics[width=0.95\textwidth]{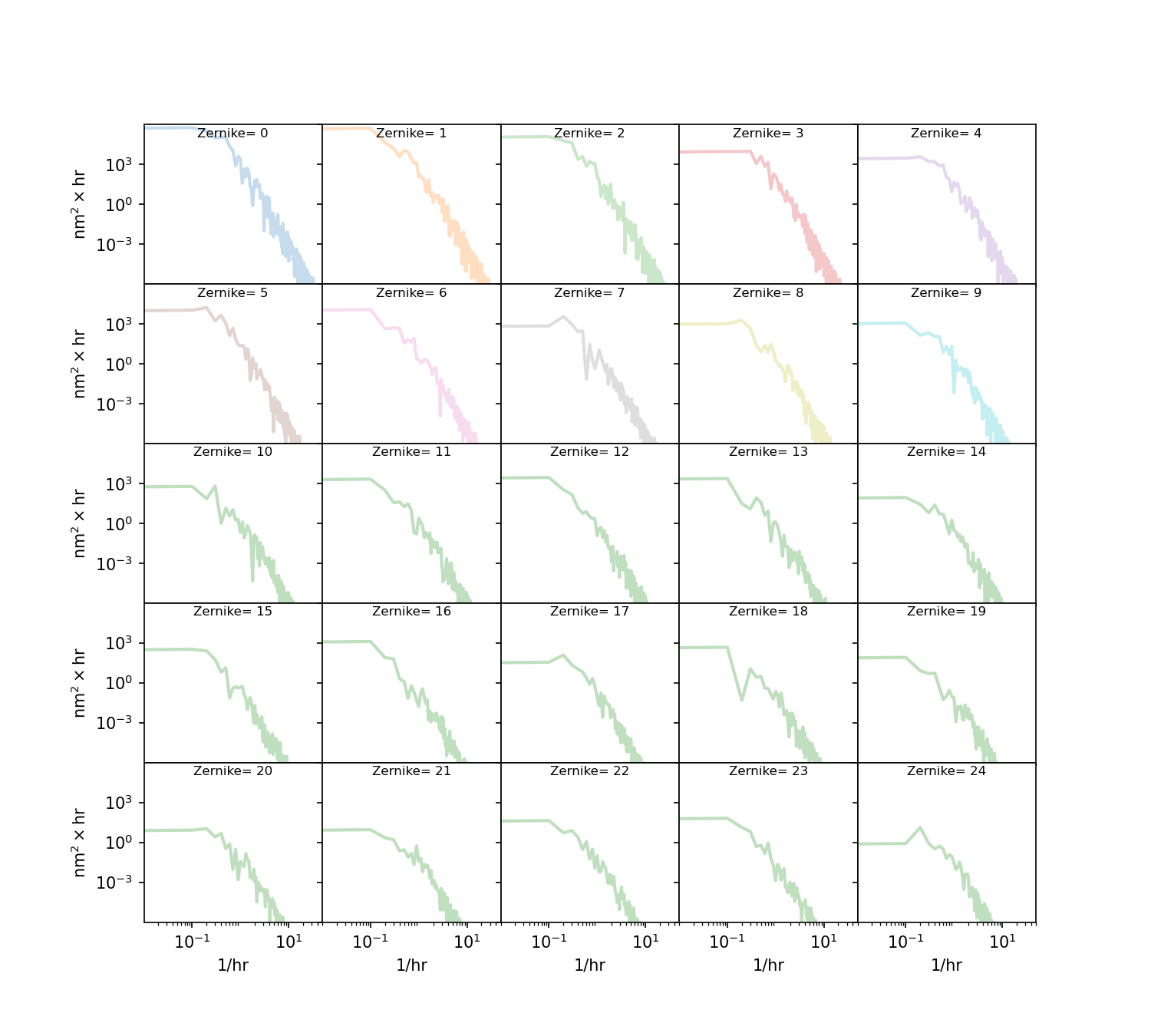}
\caption{Example realization of statistically generated spatio-temporal wavefront evolution. The top panel shows the time evolution of the ampitudes of the first few wavefront error Zernike terms, which are multiplied by Zernike coefficients to generate a 2D synthetic \gls{STOP} analysis output. The bottom grid shows the PSDs measured from time-series wavefront error amplitude functions. For these toy models, the temporal PSD has a knee at  0.45/hour and $\alpha=7$; for the spatial PSD $f_n$=5/Zernike number and $\alpha=6$, the spatial PSD falls off quickly since Seidel aberrations dominate thermal misalignment and the temporal PSD is expected to be dominated by slow thermal effects. To tolerance a system many such realizations, with different values of $\beta,\alpha$,and $f_n$ will be explored.}
\label{fig:gen_psds}
\end{center}
\end{figure}

Integrated \gls{STOP} analysis is becoming the standard for both large\cite{blaurock_structural-thermal-optical_2005,saini_impipeline_2017} and small space-telescopes\cite{ashcraft_versatile_2021}. 
These models unfortunately are often time-consuming to build and run, leading to the dilemma of whether to design the active optics system before or after the observatory design is complete and \gls{STOP} modeling of input disturbances is complete. 
To ease this causality dilemma, we adopt a parametric approach inspired by ground-based \gls{AO}\cite{males_ground-based_2018} using \glspl{PSD} early in the design process to approximate realistic disturbance and create requirements envelopes \cite{douglas_laser_2019}. 
Past work on high-contrast imaging has estimated leakage with spatio-temporal \glspl{PSD}\cite{lyon_space_2012}.
As a quick first-approximation, we generate a 2D data cube with parameterized \gls{PSD} and express the simplified von Karman \gls{PSD} as:


\begin{equation}
    PSD(f) = \frac{\beta}{(1+f/f_n)^\alpha},
\end{equation}
where $f_n$ is the so-called ``knee frequency'' of the distribution, $\alpha$ is the fall off power-law and $\beta$ is a normalizing scalar.
Fig.\,\ref{fig:gen_psds} shows the measured \gls{PSD} for an example realization of this data cube for the first 25 Zernike polynomials. 
Initial results of using this model to exercise an end-to-end observatory control system or ``digital-twin'' is described in this proceedings by Derby et al\cite{derby_integrated_2023}.
To generate a synthetic time series, each time-evolving term is assumed to start at zero and then a 40 hr. time series is generated randomly.
This results in a conservative example, since in a physical system error Zernike terms are correlated with each other and with temperature -- an opportunity for optimized predictive controllers\cite{gersh-range_improving_2014,brooks_predictive_2017,brooks_precision_2022}.
For the example plotted, the maximum gradient in focus is $\sim$1\,nm/minute, or approximately the value we expect to be able to control (see Sec. \ref{sec:active}).
This example only addresses one example \gls{PSD}; future publications will establish the bounding cases of observatory control as a function of $F_n$,   $\alpha$, and $\beta$.

\subsection{Software}
As much as possible, it is desirable to use ``off-the-shelf'' parts that have previously been space-qualified in some manner, along with designs and flight software that have been used in space or deployed to professional observatories. 
This enables significant code reuse.
 In particular, the real-time wavefront and control software inherited from the MagAO-X instrument \cite{males_magao-x:_2018}, in turned inherited from the Subaru Extreme AO system \cite{guyon_compute_2018}, is being tested on two adaptive optics testbeds at the University of Arizona Center for Astronomical Adaptive Optics\cite{van_gorkom_space_2022,schatz_three-sided_2022}. 
Ground-based observatories have long relied on the INDI  (Instrument-Neutral Distributed Interface) protocol and re-using this codebase is an appealing way to lower the cost of deploying a large space observatory. 
 Unfortunately, hardware and software faults in devices can crash INDI servers, thus the process management for MagAO-X and CAAO testbeds was recently upgraded to include a \textit{resurrector} function, which monitors for stale processes via a hexadecimal heartbeat (``hexbeat") representation of a time at some point in the future. 
If the current time exceeds the hexbeat the process is terminated and restarted. 
This code has been tested and is available publicly via the MagAO-X project\footnote{\url{https://github.com/magao-x/MagAOX/blob/resurrector/apps/resurrector/README.md}}.

\section{Managing Risks}

\subsection{Reliability}\label{sec:reliability}
Reliability comes from failure. 
Automotive manufacturers each build thousands of prototypes per year\cite{weckenborg_improving_2020} and test them on the road to discover design weaknesses.
Space missions, especially CubeSats, have a notoriously low mission success rate, $<30$\% for full mission success and $<70$\% for partial mission success, even for traditional-space companies. But the success rate increases significantly once a team has launched several missions\cite{swartwout_cubesat_2015}. 
It appears that building a complete prototype and testing it in-flight informs both the design and the engineering team's knowledge base.
A similar pattern is evident in launch vehicles, with the Space Exploration Technologies (SpaceX) Falcon 1 first succeeding on its 4th attempt\cite{clark_sweet_2008}. 
 \textit{Thus, a commitment to multiple flights is an essential part of developing a new platform or paradigm.}

Organizations vary in their ability to learn from external failures and to continue innovating despite lessons learned. 
 Academic and government NewSpace has benefited greatly from both organic and coordinated approaches to communicating lessons learned; for example the NASA Small Satellite Systems Virtual Institute  (S3VI) and its federated search tools\cite{yost_small_2018}, and the JPL \textit{F$^\prime$} (F Prime) Flight Software System\cite{bocchino_f_2018}.

Adapting SmallSat concepts to enable larger satellites at lower costs will require several ingredients:
\begin{enumerate}
\item Teams with deep knowledge of lessons already learned in both SmallSat and traditional space sectors.
\item Commitment to multiple flights to build institutional knowledge.
\item Commitment to open communication and documentation.
\end{enumerate}
The pace of the SmallSat revolution has been accelerated by openness, and while not a prerequisite for a given mission, open sharing of lessons through fora such as the CalPoly CubeSat Developers Workshop, the Utah State Small Satellite Conference as well as direct sharing of designs and source code through online repositories has enabled a burgeoning field --  as of 2023 there are over 138 NASA SmallSat science missions flown or in formulation\cite{tan_2023_2023}.

\subsection{Launch environment}
\begin{figure}
    \centering
        \includegraphics[width=.48\textwidth]{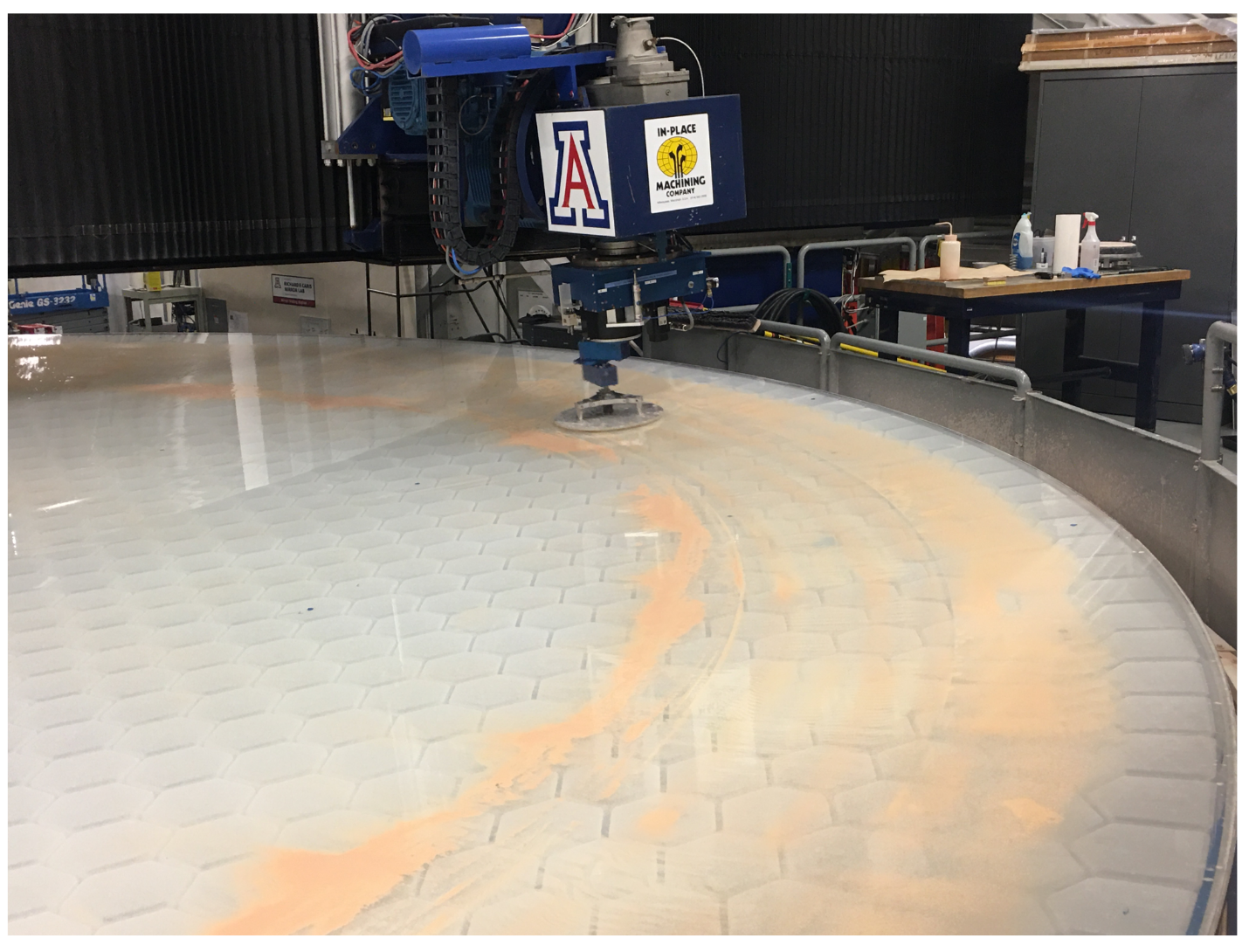}
            \includegraphics[width=.48\textwidth]{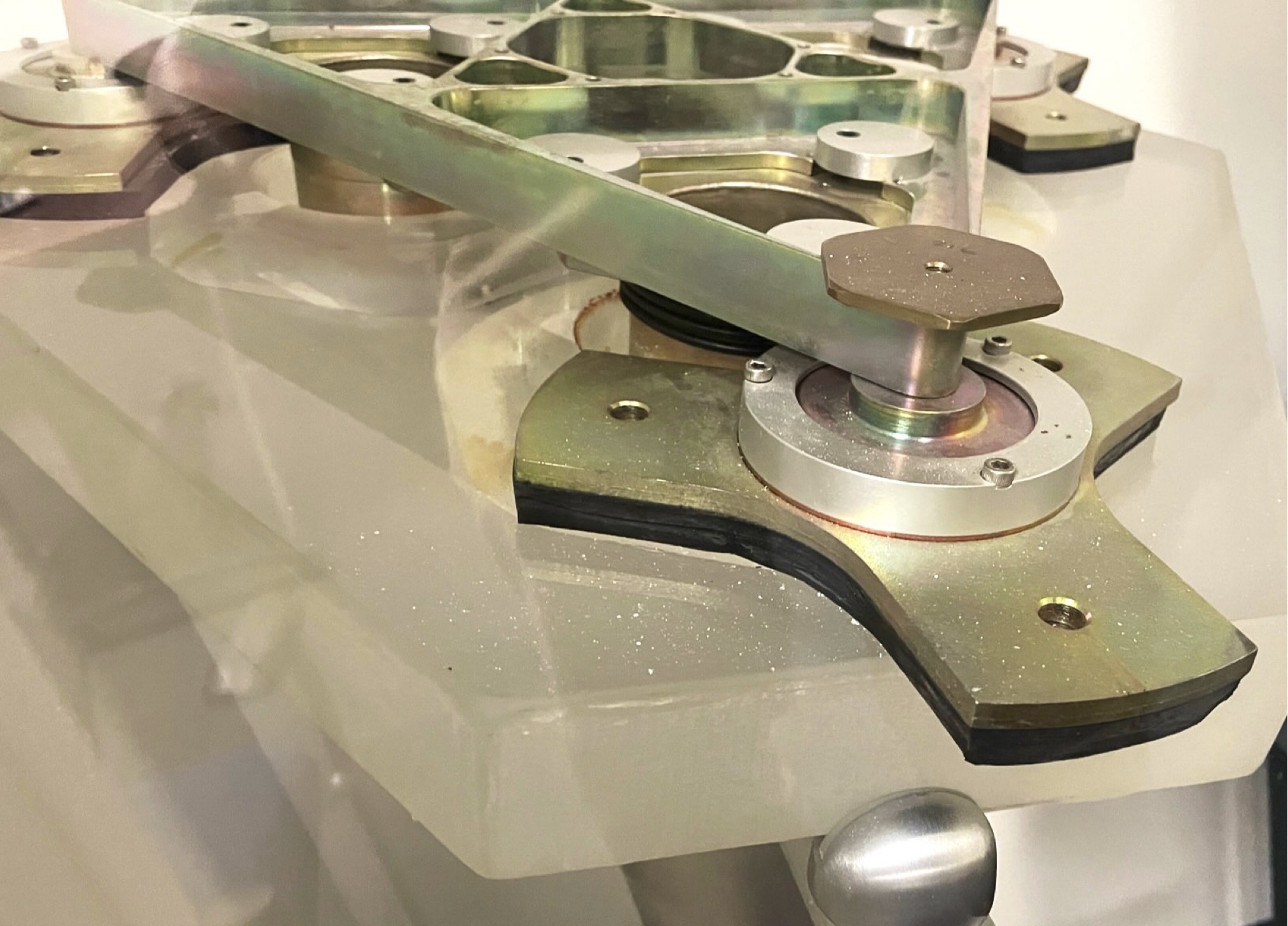}
    \caption{Left: GMT segment 3 being polished, reproduced from Martin et al\cite{martin_production_2022}. Right: Typical UA Mirror Lab load spreader mounted to a cross-section of a typical borosilicate mirror polished backside. From bottom to top: the approximate inch- thick backside, RTV bond, mounting puck, and triangular load spreader. These designs are described in detail in past papers\cite{martin_active_1998,martin_design_2006,martin_production_2012,martin_manufacture_2020,martin_production_2022}. The thick RTV layer is expected to provide significant damping, and detailed studies are in preparation that demonstrates the requirements necessary to adapt such a support system to survive launch loads (photo taken at RFCML by E.S.D.).}
    \label{fig:loadspreader}
\end{figure}

Given the large uncertainties in the expected SpaceX Starship launch environment, only preliminary enveloping studies are possible. However, UA honeycomb borosilicate mirrors (Fig.\,\ref{fig:loadspreader}) designed for the ground have several promising characteristics that suggest launch is feasible; a detailed analysis is beyond the scope of this manuscript and will be presented at a later date.

\section{Conclusion}
This manuscript provides a brief introduction to the development of space telescopes, a sampling of technologies that have enabled new concepts such as continuous wavefront control and correction of large borosilicate mirrors at lower costs, and introduces pieces of an observatory concept  that seeks to learn from successful SmallSats, ground-based observatories, as well as investments in flagship observatory technology maturation. 
Other manuscripts in these proceedings present additional related concept studies. Future work will provide details of how such an observatory might come about, additional risk mitigation strategies,  detailed instrument designs, and present the environmental requirements necessary for a honey-comb borosilicate mirror to survive launch.

\appendix    

\acknowledgments 
This manuscript is dedicated to Dr. Michael Logan Lampton\footnote{\url{https://www.ssl.berkeley.edu/ssl-mourns-the-passing-of-dr-michael-logan-lampton-1941-2023/}} and Dr. James Breckinridge\cite{pridgeon_james_2022} with deep appreciation to their contributions to the field of astronomy as a whole and these studies in particular. 

Portions of this research were supported by funding from the Technology Research Initiative Fund (TRIF) of the Arizona Board of Regents
and by generous anonymous philanthropic donations to the Steward Observatory of the College of Science at the University of Arizona and to the Space Science Laboratory of the University of California, Berkeley. 

\bibliography{report,esd-zotero} 
\bibliographystyle{spiebib} 

\end{document}

%% file: acronyms.tex


\newacronym{AU}{AU}{astronomical Unit [1.5e11 m]}  
\newacronym{pc}{pc}{parsec}
\newacronym{mas}{mas}{milliarcsecond}
\newacronym{nm}{nm}{nanometer}
\newacronym{CTE}{CTE}{coefficient of thermal expansion}
\newacronym{sqarc}{$as^2$}{square arcsecond}

\newacronym{smc}{SMC}{Small Magellanic Cloud}
\newacronym{lmc}{LMC}{Large Magellanic Cloud}
\newacronym{ism}{ISM}{interstellar medium}
\newacronym{mw}{MW}{Milky Way}
\newacronym{epseri}{$\epsilon$ Eri}{Epsilon Eridani}
\newacronym{EKB}{EKB}{Edgeworth-Kuiper Belt}

\newacronym{CFR}{CFR}{Complete Frequency Redistribution}

\newacronym{nasa}{NASA}{National Aeronautics and Space Agency}
\newacronym{esa}{ESA}{European Space Agency}
\newacronym{omi}{OMI}{\textit{Optical Mechanics Inc.}}
\newacronym{gsfc}{GSFC}{\gls{nasa} Goddard Space Flight Center}
\newacronym{stsci}{STScI}{Space Telescope Science Institute}
\newacronym{nsroc}{NSROC}{\gls{nasa} Sounding Rocket Operations Contract}
\newacronym{wff}{WFF}{\gls{nasa} Wallops Flight Facility}
\newacronym{wsmr}{WSMR}{White Sands Missile Range}

\newacronym{irac}{IRAC}{Infrared Array Camera}
\newacronym[plural=CCDs, firstplural=charge-coupled devices (CCDs)]{ccd}{CCD}{charge-coupled device}
\newacronym[plural=EMCCDs, firstplural=electron multiplying charge-coupled devices (EMCCDs)]{EMCCD}{EMCCD}{electron multiplying charge-coupled device}

\newacronym{DM}{DM}{Deformable Mirror}
\newacronym{MCP}{MCP}{ Microchannel Plate }
\newacronym{ipc}{IPC}{Image Proportional Counter}
\newacronym{cots}{COTS}{Commercial Off-The-Shelf}
\newacronym{ISR}{ISR}{incoherent scatter radar}
\newacronym{atcamera}{AT}{angle tracker}
\newacronym{MEMS}{MEMS}{microelectromechanical systems}
\newacronym{QE}{QE}{quantum efficiency}
\newacronym{RTD}{RTD}{Resistance Temperature Detector}
\newacronym{PID}{PID}{Proportional-Integral-Derivative}
\newacronym{PRNU}{PRNU}{photo response non-uniformity}
\newacronym{DSNU}{PRNU}{dark signal non-uniformity}
\newacronym{CMOS}{CMOS}{complementary metal–oxide–semiconductor}
\newacronym{TRL}{TRL}{technology readiness level}
\newacronym{swap}{SWaP}{Size, Weight, and Power}
\newacronym{ConOps}{ConOps}{concept of operations}
\newacronym{NRE}{NRE}{non-recurring engineering}
\newacronym{CBE}{CBE}{current best estimate}

\newacronym{FOV}{FOV}{field-of-view}
\newacronym{NIR}{NIR}{near-infrared}
\newacronym{PV}{PV}{Peak-to-Valley}
\newacronym{MRF}{MRF}{Magnetorheological finishing}
\newacronym{AO}{AO}{Adaptive Optics}
\newacronym{TTP}{TTP}{tip, tilt, and piston}
\newacronym{FPS}{FPS}{fine pointing system}
\newacronym{SHWFS}{SHWFS}{Shack-Hartmann Wavefront Sensor}
\newacronym{OAP}{OAP}{off-axis parabola}
\newacronym{LGS}{LGS}{laser guide star}
\newacronym{WFCS}{WFCS}{wavefront control system}
\newacronym{OPD}{OPD}{optical path difference}
\newacronym{UA}{UA}{University of Arizona}
\newacronym{MEL}{MEL}{Master Equipment List}
\newacronym{LEO}{LEO}{low-earth orbit}
\newacronym{GEO}{GEO}{geosynchronous orbit}
\newacronym{EFC}{EFC}{electric-field conjugation}
\newacronym{LDFC}{LDFC}{linear dark field control}
\newacronym{DAC}{DAC}{digital-to-analog converter}
\newacronym{FEA}{FEA}{finite element analysis}

\newacronym{SiC}{SiC}{Silicon Carbide}
\newacronym{ESPA}{ESPA}{EELV Secondary Payload Adapter}
\newacronym{EEID}{EEID}{Earth-equivalent Insolation Distance, the distance from the star where the incident energy density is that of the Earth received from the Sun}
\newacronym{LLOWFS}{LLOWFS}{Lyot low-order wavefront sensor}
\newacronym{STOP}{STOP}{Structural-Thermal-Optical-Performance}

\newacronym{resel}{resel}{resolution element}

\newacronym{acs}{ACS}{Attitude Control System}
\newacronym{orsa}{ORSA}{Ogive Recovery System Assembly}
\newacronym{gse}{GSE}{Ground Station Equipment}
\newacronym{FSM}{FSM}{Fast Steering Mirror}

\newacronym{WFS}{WFS}{wavefront sensor}
\newacronym{LSI}{LSI}{Lateral Shearing Interferometer}
\newacronym{VVC}{VVC}{Vector Vortex Coronagraph}
\newacronym{VNC}{VNC}{Visible Nulling Coronagraph}
\newacronym{CGI}{CGI}{Coronagraph Instrument}
\newacronym{IWA}{IWA}{Inner Working Angle}
\newacronym{OWA}{OWA}{Outer Working Angle}
\newacronym{NPZT}{N-PZT}{Nuller piezoelectric transducer}
\newacronym{ZWFS}{ZWFS}{Zernike wavefront sensor}
\newacronym{SPC}{SPC}{Shaped Pupil Coronagraph}
\newacronym{HLC}{HLC}{Hybrid-Lyot Coronagraph}
\newacronym{ADI}{ADI}{angular differential imaging}
\newacronym{RDI}{RDI}{reference differential imaging}
\newacronym{LOWFSC}{LOWFS/C}{low-order wavefront sensing and control}
\newacronym{HOWFSC}{HOWFS/C}{high-order wavefront sensing and control}
\newacronym{WFSC}{WFSC}{wavefront sensing and control}

\newacronym{HST}{HST}{Hubble Space Telescope}
\newacronym{GPS}{GPS}{Global Positioning System}
\newacronym{ISS}{ISS}{International Space Station}
\newacronym[description=Advanced CCD Imaging Spectrometer]{acis}{ACIS}{Advanced \gls{ccd} Imaging Spectrometer}
\newacronym{stis}{STIS}{\textit{Space Telescope Imaging Spectrograph}}
\newacronym{mcp}{MCP}{Microchannel Plate}
\newacronym{jwst}{JWST}{$\textit{JWST}$}
\newacronym{fuse}{FUSE}{$\textit{FUSE}$}
\newacronym{galex}{GALEX}{$\textit{Galaxy Evolution Explorer}$}
\newacronym{spitzer}{Spitzer}{$\textit{Spitzer Space Telescope}$}
\newacronym{mips}{MIPS}{Multiband Imaging Photometer for \gls{spitzer}}
\newacronym{gissmo}{GISSMO}{Gas Ionization Solar Spectral Monitor}
\newacronym{iue}{IUE}{International Ultraviolet Explorer}
\newacronym{spinr}{SPINR}{$\textit{Spectrograph for Photometric Imaging with Numeric Reconstruction}$}
\newacronym{imager}{IMAGER}{$\textit{Interstellar Medium Absorption Gradient Experiment Rocket}$}
\newacronym{TPF-C}{TPF-C}{Terrestrial Planet Finder Coronagraph}
\newacronym{RAIDS}{RAIDS}{Atmospheric and Ionospheric Detection System }
\newacronym{mama}{MAMA}{Multi-Anode Microchannel Array}
\newacronym{ATLAST}{ATLAST}{Advanced Technology Large Aperture Space Telescope}
\newacronym{PICTURE}{PICTURE}{Planet Imaging Concept Testbed Using a Rocket Experiment}
\newacronym{LITES}{LITES}{Limb-imaging Ionospheric and Thermospheric
Extreme-ultraviolet Spectrograph}
\newacronym{LBT}{LBT}{Large Binocular Telescope}
\newacronym{LBTI}{LBTI}{Large Binocular Telescope Interferometer}
\newacronym{KIN}{KIN}{Keck Interferometer Nuller}
\newacronym{SHARPI}{SHARPI}{Solar High-Angular Resolution Photometric Imager}
\newacronym{IRAS}{IRAS}{Infrared Astronomical Satellite}
\newacronym{HARPS}{HARPS}{High Accuracy Radial velocity Planetary}
\newacronym{hstSTIS}{STIS}{Space Telescope Imaging Spectrograph}
\newacronym{spitzerIRAC}{IRAC}{Infrared Array Camera}
\newacronym{spitzerMIPS}{MIPS}{Multiband Imaging Photometer for Spitzer}
\newacronym{spitzerIRS}{IRS}{Infrared Spectrograph}
\newacronym{CHARA}{CHARA}{Center for High Angular Resolution Astronomy}
\newacronym{wfirst-afta}{WFIRST-AFTA}{Wide-Field InfrarRed Survey
Telescope-Astrophysics Focused Telescope Assets}
\newacronym{GPI}{GPI}{Gemini Planet Imager}
\newacronym{WFIRST}{Roman}{Nancy Grace Roman Space Telescope}
\newacronym{HabEx}{HabEx}{Habitable Exoplanet Observatory Mission Concept}
\newacronym{LUVOIR}{LUVOIR}{Large UV/Optical/Infrared Surveyor}
\newacronym{FGS}{FGS}{Fine Guidance Sensor}
\newacronym{STIS}{STIS}{Space Telescope Imaging Spectrograph}
\newacronym{MGHPCC}{MGHPCC}{Massachusetts Green High Performance
Computing Center}
\newacronym{WISE}{WISE}{Wide-field Infrared Survey Explorer}
\newacronym{ALMA}{ALMA}{Atacama Large Millimeter Array}
\newacronym{GRAIL}{GRAIL}{Gravity Recovery and Interior Laboratory}
\newacronym{jwstNIRCam}{NIRCam}{near-\gls{IR}-camera}
\newacronym{jwstMIRI}{MIRI}{Mid-Infrared Instrument}

\newacronym{AURIC}{AURIC}{The Atmospheric Ultraviolet Radiance Integrated Code} 
\newacronym{FFT}{FFT}{Fast Fourier Transform  }
\newacronym{MODTRAN}{MODTRAN   }{ MODerate resolution atmospheric TRANsmission }
\newacronym{idl}{IDL}{$\textit {Interactive Data Language}$}
\newacronym[sort=NED,description=NASA/IPAC Extragalactic Database]{ned}{NED}{\gls{nasa}/\gls{ipac} Extragalactic Database}
\newacronym{iraf}{IRAF}{Image Reduction and Analysis Facility}
\newacronym{wcs}{WCS}{World Coordinate System}
\newacronym{pegase}{PEGASE}{$\textit{Projet d'Etude des GAlaxies par Synthese Evolutive}$}
\newacronym{dirty}{DIRTY}{$\textit{DustI Radiative Transfer, Yeah!}$}
\newacronym{CUDA}{CUDA}{Compute Unified Device Architecture}
\newacronym{KLIP}{KLIP}{Karhunen-Lo`eve Image Processing}
\newacronym{FEM}{FEM}{finite element method}

\newacronym{MSIS}{MSIS}{Mass Spectrometer Incoherent Scatter Radar}
\newacronym{nmf2}{$N_m$}{F2-Region Peak density}
\newacronym{hmf2}{$h_m$}{F2-Region Peak height}
\newacronym{H}{$H$}{F2-Region Scale Height}

\newacronym{isr}{ISR}{Incoherent Scatter Radar}
\newacronym[description=TLA Within Another Acronym]{twaa}{TWAA}{\gls{tla} Within Another Acronym}
\newacronym[plural=SNe, firstplural=Supernovae (SNe)]{sn}{SN}{Supernova}
\newacronym{EUV}{EUV}{Extreme-Ultraviolet }
\newacronym{EUVS}{EUVS}{\gls{EUV} Spectrograph}
\newacronym{F2}{F2}{Ionospheric Chapman F Layer }
\newacronym{F10.7}{F10.7}{ 10.7 cm radio flux [10$^{-22}$ W m$^{-2}$ Hz$^{-1}$]  }
\newacronym{FUV}{FUV}{far-ultraviolet }
\newacronym{IR}{IR}{infrared}
\newacronym{MUV}{MUV}{mid-ultraviolet }
\newacronym{NUV}{NUV}{near-ultraviolet }
\newacronym{O$^+$}{O$^+$}{Singly Ionized Oxygen  Atom }
\newacronym{OI}{OI}{Neutral Atomic Oxygen Spectroscopic State }
\newacronym{OII}{OII}{Singly Ionized Atomic Oxygen Spectroscopic State }
\newacronym{PSF}{PSF}{point spread function}
\newacronym{$R_E$}{$R_E$}{Earth radii [$\approx$ 6400 km]  }
\newacronym{RV}{RV}{radial velocity}
\newacronym{UV}{UV}{ultraviolet }
\newacronym{WFE}{WFE}{wavefront error}
\newacronym{sed}{SED}{spectral energy distribution}
\newacronym{nir}{NIR}{near-infrared}
\newacronym{mir}{MIR}{mid-infrared}
\newacronym{ir}{IR}{infrared}
\newacronym{uv}{UV}{ultraviolet}
\newacronym[plural=PAHs, firstplural=Polycyclic Aromatic Hydrocarbons (PAHs)]{pah}{PAH}{Polycyclic Aromatic Hydrocarbon}
\newacronym{obsid}{OBSID}{Observation Identification}
\newacronym{SZA}{SZA}{Solar Zenith Angle}
\newacronym{TLE}{TLE}{Two Line Element set}
\newacronym{DOF}{DOF}{degrees-of-freedom}
\newacronym{PZT}{PZT}{lead zirconate titanate}
\newacronym{ADCS}{ADCS}{attitude determination and control system}
\newacronym{COTS}{COTS}{commercial off-the-shelf}
\newacronym{CDH}{C$\&$DH}{command and data handling}
\newacronym{EPS}{EPS}{electrical power system}

\newacronym{PCA}{PCA}{principal component analysis}
\newacronym{fwhm}{FWHM}{full-width-half maximum}
\newacronym{RMS}{RMS}{root mean squared}
\newacronym{RMSE}{RMSE}{root mean squared error}
\newacronym{MCMC}{MCMC}{Marcov chain Monte Carlo}
\newacronym{DIT}{DIT}{discrete inverse theory}
\newacronym{SNR}{SNR}{signal-to-noise ratio}
\newacronym{PSD}{PSD}{power spectral density}
\newacronym{NMF}{NMF}{non-negative matrix factorization}

%% file: main.bbl
\begin{thebibliography}{100}

\bibitem{roman_lst_1974}
Roman, N.~G., ``{{LST}}: {{The National Space Observatory}} concept - {{An}}
  observatory for future astronomer involvement in planning and use,'' in [{\em
  {{NTRS Meeting Information}}: {{Aerospace Sciences Meeting}}; 1974-01-30 to
  1974-02-01}{\nolinebreak\hspace{0.1em}]},  {NTRS Document ID: 19740056405},
  {Washington, D.C.} (Jan. 1974).

\bibitem{linsky_uv_2018}
Linsky, J.~L., ``{{UV}} astronomy throughout the ages: a historical
  perspective,'' {\em Astrophys Space Sci}~{\bf 363},  101 (Apr. 2018).

\bibitem{rieke_last_2021}
Rieke, G.~H.,  [{\em The {{Last}} of the {{Great Observatories}}: {{Spitzer}}
  and the {{Era}} of {{Faster}}, {{Better}}, {{Cheaper}} at
  {{NASA}}}{\nolinebreak\hspace{0.1em}]}, {University of Arizona Press} (Nov.
  2021).

\bibitem{smith_nasa_2018}
Smith, L., Lucas, F.~D., Rohrabacher, D., Brooks, M., Hultgren, R., Posey, B.,
  Massie, T., Weber, R.~K., Knight, S., Babin, B., Comstock, B., Loudermilk,
  B., Abraham, R.~L., Palmer, G., Webster, D., Banks, J., Biggs, A., Marshall,
  R.~W., Dunn, N.~P., Higgins, C., Norman, R., Carolina, S., Lesko, D.,
  Johnson, E.~B., Lofgren, Z., Lipinski, D., Bonamici, S., Bera, A., Esty,
  E.~H., Veasey, M.~A., Beyer, D.~S., Rosen, J., Lamb, C., and Babin, H.~B.,
  ``{{NASA}} cost and schedule overruns: acquisition and program management
  challenges,'' (June 2018).

\bibitem{quintana_pandora_2021}
Quintana, E.~V., Col{\'o}n, K.~D., Mosby, G., Schlieder, J.~E., Supsinskas, P.,
  Karburn, J., Dotson, J.~L., Greene, T.~P., Hedges, C., Apai, D., Barclay, T.,
  Christiansen, J.~L., Espinoza, N., Mullally, S.~E., Gilbert, E.~A., Hoffman,
  K., Kostov, V.~B., Lewis, N.~K., Foote, T.~O., Mason, J., Youngblood, A.,
  Morris, B.~M., Newton, E.~R., Pepper, J., Rackham, B.~V., Rowe, J.~F., and
  Stevenson, K., ``The {{Pandora SmallSat}}: {{Multiwavelength
  Characterization}} of {{Exoplanets}} and their {{Host Stars}},'' (Aug. 2021).

\bibitem{2004SPIE.5489..563V}
{van Belle}, G.~T., {Meinel}, A.~B., and {Meinel}, M.~P., ``{The scaling
  relationship between telescope cost and aperture size for very large
  telescopes},'' in [{\em Ground-based
  Telescopes}{\nolinebreak\hspace{0.1em}]},  {Oschmann}, Jacobus~M., J., ed.,
  {\em Society of Photo-Optical Instrumentation Engineers (SPIE) Conference
  Series} {\bf 5489},  563--570 (Oct. 2004).

\bibitem{2021Natur.600..208W}
{Witze}, A., ``{The \$11-billion Webb telescope aims to probe the early
  Universe},'' {\em Nature}~{\bf 600},  208--212 (Dec. 2021).

\bibitem{jwst_icrp}
Ballhaus, W.~F., Casani, J., Dorman, S., Gallagher, D., Illingworth, G.,
  Klineberg, J., Schurr, D., Lewis, R., and Lobbia, M., ``James webb space
  telescope independent comprehensive review panel final report,'' tech. rep.,
  JWST-ICRP (2010).

\bibitem{elvis_accelerating_2023}
Elvis, M., Lawrence, C., and Seager, S., ``Accelerating astrophysics with the
  {{SpaceX Starship}},'' {\em Physics Today}~{\bf 76},  40--45 (Feb. 2023).

\bibitem{angel_manufacture_1982}
Angel, J. R.~P. and Hill, J.~M., ``Manufacture {{Of Large Glass Honeycomb
  Mirrors}},'' in [{\em Advanced {{Technology Optical Telescopes
  I}}}{\nolinebreak\hspace{0.1em}]},   {\bf 0332},  298--306, {SPIE} (Nov.
  1982).

\bibitem{nelson_construction_1988}
Nelson, J.~E. and Mast, T.~S., ``Construction of the {{Keck Observatory}},'' in
  [{\em Proceedings of a {{ESO Conference}} on {{Very Large Telescopes}} and
  their {{Instrumentation}}}{\nolinebreak\hspace{0.1em}]},   {\bf 30},  7 (Oct.
  1988).

\bibitem{martin_newspace_2017}
Martin, G., ``{{NewSpace}}: {{The Emerging Commercial Space Industry}},'' in
  [{\em {{ISU Master Class Lecture}}}{\nolinebreak\hspace{0.1em}]},  (Feb.
  2017).

\bibitem{rizk_ocams_2018}
Rizk, B., {Drouet~d'Aubigny}, C., Golish, D., Fellows, C., Merrill, C., Smith,
  P., Walker, M.~S., Hendershot, J.~E., Hancock, J., Bailey, S.~H.,
  DellaGiustina, D.~N., Lauretta, D.~S., Tanner, R., Williams, M., Harshman,
  K., Fitzgibbon, M., Verts, W., Chen, J., Connors, T., Hamara, D., Dowd, A.,
  Lowman, A., Dubin, M., Burt, R., Whiteley, M., Watson, M., McMahon, T., Ward,
  M., Booher, D., Read, M., Williams, B., Hunten, M., Little, E., Saltzman, T.,
  Alfred, D., O'Dougherty, S., Walthall, M., Kenagy, K., Peterson, S.,
  Crowther, B., Perry, M.~L., See, C., Selznick, S., Sauve, C., Beiser, M.,
  Black, W., Pfisterer, R.~N., Lancaster, A., Oliver, S., Oquest, C., Crowley,
  D., Morgan, C., Castle, C., Dominguez, R., and Sullivan, M., ``{{OCAMS}}:
  {{The OSIRIS-REx Camera Suite}},'' {\em Space Sci Rev}~{\bf 214},  26 (Jan.
  2018).

\bibitem{chung_aspera_2021}
Chung, H., Vargas, C.~J., Hamden, E., McMahon, T., Gonzales, K., Khan, A.~R.,
  Agarwal, S., Bailey, H., Behroozi, P., Brendel, T., Choi, H., Connors, T.,
  Corlies, L., Corliss, J., Dettmar, R.-J., Dolana, D., Douglas, E.~S., Guzman,
  J., Hamara, D., Harris, W., Harshman, K., Hergenrother, C., Hoadley, K.,
  Kidd, J., Kim, D., Li, J.~S., Montoya, M., Sauve, C., Schiminovich, D.,
  Selznick, S., Siegmund, O., Ward, M., Wolcott, E.~M., and Zaritsky, D.,
  ``Aspera: the {{UV SmallSat}} telescope to detect and map the warm-hot gas
  phase in nearby galaxy halos,'' in [{\em {{UV}}/{{Optical}}/{{IR Space
  Telescopes}} and {{Instruments}}: {{Innovative Technologies}} and {{Concepts
  X}}}{\nolinebreak\hspace{0.1em}]},   {\bf 11819},  1181903, {International
  Society for Optics and Photonics} (Aug. 2021).

\bibitem{spitzer_space_1960}
Spitzer, L., ``Space telescopes and components,'' {\em The Astronomical
  Journal}~{\bf 65},  242 (June 1960).

\bibitem{spitzer_beginnings_1962}
SPITZER, {\relax LYMAN}., ``{{THE BEGINNINGS AND FUTURE OF SPACE ASTRONOMY}},''
  {\em American Scientist}~{\bf 50}(3),  473--484 (1962).

\bibitem{roman_aip_1980}
Roman, N.~G. and DeVorkin, D., ``{{AIP ORAL HISTORIES Nancy G}}. {{Roman}},''
  (Aug. 1980).

\bibitem{hammond_ultraviolet_1970}
Hammond, A.~L., ``Ultraviolet {{Astronomy}}: {{Progress}} with the {{OAO}},''
  {\em Science}~{\bf 170},  960--961 (Nov. 1970).

\bibitem{shockey_study_1981}
Shockey, E.~F., ``A study of the longevity and operational reliability of
  {{Goddard Spacecraft}}, 1960-1980,'' Tech. Rep. NASA-TM-82178 (Aug. 1981).

\bibitem{aumann_infrared_1977}
Aumann, H.~H. and Walker, R.~G., ``Infrared {{Astronomical Satellite}},'' {\em
  Opt. Eng}~{\bf 16}(6),  166537--166537-- (1977).

\bibitem{boggess_iue_1978}
Boggess, A., Carr, F.~A., Evans, D.~C., Fischel, D., Freeman, H.~R., Fuechsel,
  C.~F., Klinglesmith, D.~A., Krueger, V.~L., Longanecker, G.~W., Moore, J.~V.,
  Pyle, E.~J., Rebar, F., Sizemore, K.~O., Sparks, W., Underhill, A.~B.,
  Vitagliano, H.~D., West, D.~K., Macchetto, F., Fitton, B., Barker, P.~J.,
  Dunford, E., Gondhalekar, P.~M., Hall, J.~E., Harrison, V. a.~W., Oliver,
  M.~B., Sandford, M. C.~W., Vaughan, P.~A., Ward, A.~K., Anderson, B.~E.,
  Boksenberg, A., Coleman, C.~I., Snijders, M. a.~J., and Wilson, R., ``The
  {{IUE}} spacecraft and instrumentation,'' {\em Nature}~{\bf 275},  372--377
  (Oct. 1978).

\bibitem{meinel_high_1962}
Meinel, A.~B.,  [{\em High {{Resolution Optical Space
  Telescopes}}}{\nolinebreak\hspace{0.1em}]} (Jan. 1962).

\bibitem{meinel_new_1961}
Meinel, A.~B., ``New {{Frontiers}} of {{Astronomical Technology}}:
  {{Technological}} developments challenge the astronomer, both from the ground
  and in space.,'' {\em Science}~{\bf 134},  1165--1171 (Oct. 1961).

\bibitem{richard_dynamic_1988}
Richard, R.~M., Cho, M., and Pollard, W., ``Dynamic {{Analysis Of The SIRTF
  One-Meter Mirror During Launch}},'' in [{\em Cryogenic {{Optical Systems}}
  and {{Instruments III}}}{\nolinebreak\hspace{0.1em}]},   {\bf 0973},  86--99,
  {SPIE} (Apr. 1988).

\bibitem{baiocchi_demonstration_2001}
Baiocchi, D., Burge, J.~H., and Cuerden, B., ``Demonstration of a 0.5-m
  ultralightweight mirror for use at geosynchronous orbit,'' in [{\em Optical
  {{Manufacturing}} and {{Testing IV}}}{\nolinebreak\hspace{0.1em}]},   {\bf
  4451},  86--95, {SPIE} (Dec. 2001).

\bibitem{baiocchi_design_2004}
Baiocchi, D., {\em Design and control of lightweight, active space mirrors},
  PhD thesis (Dec. 2004).

\bibitem{baiocchi_optimized_2004}
Baiocchi, D. and Burge, J.~H., ``Optimized active lightweight space mirrors,''
  in [{\em {{UV}}/{{Optical}}/{{IR Space Telescopes}}: {{Innovative
  Technologies}} and {{Concepts}}}{\nolinebreak\hspace{0.1em}]},   {\bf 5166},
  49--57, {SPIE} (Jan. 2004).

\bibitem{cohan_vibroacoustic_2011}
Cohan, L.~E. and Miller, D.~W., ``Vibroacoustic launch analysis and alleviation
  of lightweight, active mirrors,'' {\em OE}~{\bf 50},  013002 (Jan. 2011).

\bibitem{saif_nanometer_2015}
Saif, B., Chaney, D., Smith, W.~S., Greenfield, P., Hack, W., Bluth, J., Otten,
  A.~V., Bluth, M., Sanders, J., {Keski-Kuha}, R., Feinberg, L.,
  {North-Morris}, M., and Millerd, J., ``Nanometer level characterization of
  the {{James Webb Space Telescope}} optomechanical systems using high-speed
  interferometry,'' {\em Appl. Opt., AO}~{\bf 54},  4285--4298 (May 2015).

\bibitem{briguglio_development_2017}
Briguglio, R., Xompero, M., Riccardi, A., Lisi, F., Du{\`o}, F., Vettore, C.,
  Gallieni, D., Tintori, M., Lazzarini, P., Patauner, C., Biasi, R., D'Amato,
  F., Pucci, M., and do~Carmo, J.~P., ``Development of large aperture telescope
  technology ({{LATT}}): test results on a demonstrator bread-board,'' in [{\em
  International {{Conference}} on {{Space Optics}} \textemdash{} {{ICSO}}
  2014}{\nolinebreak\hspace{0.1em}]},   {\bf 10563},  1275--1283, {SPIE} (Nov.
  2017).

\bibitem{junkins_near-minimum-time_1991}
Junkins, J.~L., Rahman, Z.~H., and Bang, H., ``Near-minimum-time control of
  distributed parameter systems - {{Analytical}} and experimental results,''
  {\em Journal of Guidance, Control, and Dynamics}~{\bf 14},  406--415 (Mar.
  1991).

\bibitem{saenz-otero_using_2005}
{Saenz-Otero}, A. and Miller, D.~W., ``Using {{ISS}} to develop telescope
  technology,'' in [{\em {{UV}}/{{Optical}}/{{IR Space Telescopes}}:
  {{Innovative Technologies}} and {{Concepts
  II}}}{\nolinebreak\hspace{0.1em}]},   {\bf 5899},  172--183, {SPIE} (Aug.
  2005).

\bibitem{crawley_middeck_1993}
Crawley, E.~F., Vanschoor, M.~C., and Bokhour, E.~B., ``The middeck 0-gravity
  dynamics experiment,'' Tech. Rep. NASA-CR-4500 (Jan. 1993).

\bibitem{miller_middeck_1995}
Miller, D., {de Luis}, J., Stover, G., How, J., Liu, K., Grocott, S., Campbell,
  M., Glaese, R., and Crawley, E., ``The {{Middeck Active Control Experiment}}
  ({{MACE}}): using space for technology research and development,'' in [{\em
  Proceedings of 1995 {{American Control Conference}} -
  {{ACC}}'95}{\nolinebreak\hspace{0.1em}]},   {\bf 1},  397--401 vol.1 (June
  1995).

\bibitem{brooks_advanced_2015-1}
Brooks, T., Stahl, H.~P., and Sr, W. R.~A., ``Advanced {{Mirror Technology
  Development}} ({{AMTD}}) thermal trade studies,'' in [{\em Optical
  {{Modeling}} and {{Performance Predictions
  VII}}}{\nolinebreak\hspace{0.1em}]},   {\bf 9577},  957703, {SPIE} (Sept.
  2015).

\bibitem{brooks_precision_2022}
Brooks, T.~E. and Stahl, H.~P., ``Precision thermal control technology to
  enable thermally stable telescopes,'' {\em JATIS}~{\bf 8},  024001 (Apr.
  2022).

\bibitem{stahl_jwst_2004}
Stahl, H.~P., Feinberg, L.~D., and Texter, S.~C., ``{{JWST}} primary mirror
  material selection,'' in [{\em Proc. {{SPIE}}}{\nolinebreak\hspace{0.1em}]},
   {\bf 5487},  818--825, {International Society for Optics and Photonics}
  (Oct. 2004).

\bibitem{stahl_jwst_2007}
Stahl, H.~P., ``{{JWST}} mirror technology development results,'' in [{\em
  Optical {{Manufacturing}} and {{Testing VII}}}{\nolinebreak\hspace{0.1em}]},
   {\bf 6671},  11--22, {SPIE} (Sept. 2007).

\bibitem{matthews_kodak_2003}
Matthews, G., Barrett, D., Bolton, J., Dahl, R., Michaels, E., Mallette, M.,
  and Johnson, J., ``Kodak {{AMSD}} mirror program: overview and cryo test
  results,'' in [{\em Optical {{Manufacturing}} and {{Testing
  V}}}{\nolinebreak\hspace{0.1em}]},   {\bf 5180},  169--179, {SPIE} (Dec.
  2003).

\bibitem{catanzaro_herschel_2009}
Catanzaro, B. and Doyle, D., ``Herschel {{Space Telescope}}: {{Optical}} test
  and model correlation,'' in [{\em 2009 {{IEEE Aerospace}}
  conference}{\nolinebreak\hspace{0.1em}]},   1--14 (Mar. 2009).

\bibitem{west_space_2010}
West, S.~C., Bailey, S.~H., Bauman, S., Cuerden, B., Granger, Z., and Olbert,
  B.~H., ``A space imaging concept based on a 4m structured spun-cast
  borosilicate monolithic primary mirror,'' in [{\em {{SPIE Astronomical
  Telescopes}} + {{Instrumentation}}}{\nolinebreak\hspace{0.1em}]},  Oschmann,
  Jr., J.~M., Clampin, M.~C., and MacEwen, H.~A., eds.,  77311O (July 2010).

\bibitem{eads_20_2020-1}
Eads, R.~W. and Angel, J. R.~P., ``A 20 m wide-field diffraction-limited
  telescope,'' {\em Philosophical Transactions of the Royal Society A:
  Mathematical, Physical and Engineering Sciences}~{\bf 379},  20200141 (Nov.
  2020).

\bibitem{schneider_owl-moon_2022}
Schneider, J., Silk, J., and Vakili, F., ``{{OWL-Moon}}: {{Very}} high
  resolution spectropolarimetric interferometry and imaging from the {{Moon}}:
  exoplanets to cosmology,'' {\em Exp Astron}  (Aug. 2022).

\bibitem{martin_next-generation_2022}
Martin, S.~R., Lawrence, C.~R., Redding, D.~C., Mennesson, B., Rodgers, J.~M.,
  Hurd, K., Morgan, R.~M., Hu, R., Steeves, J.~B., Jewell, J.~B., Phillips, C.,
  Pineda, C., Ferraro, N., and Flinois, T. L.~B., ``Next-generation active
  telescope for space astronomy,'' {\em JATIS}~{\bf 8},  044005 (Dec. 2022).

\bibitem{marlow_laser-guide-star_2017}
Marlow, W.~A., Carlton, A.~K., Yoon, H., Clark, J.~R., Haughwout, C.~A., Cahoy,
  K.~L., Males, J.~R., Close, L.~M., and Morzinski, K.~M., ``Laser-{{Guide-Star
  Satellite}} for {{Ground-Based Adaptive Optics Imaging}} of {{Geosynchronous
  Satellites}},'' {\em Journal of Spacecraft and Rockets}~{\bf 54}(3),
  621--639 (2017).

\bibitem{douglas_laser_2019}
Douglas, E.~S., Males, J.~R., Clark, J., Guyon, O., Lumbres, J., Marlow, W.,
  and Cahoy, K.~L., ``Laser {{Guide Star}} for {{Large Segmented-aperture Space
  Telescopes}}. {{I}}. {{Implications}} for {{Terrestrial Exoplanet Detection}}
  and {{Observatory Stability}},'' {\em AJ}~{\bf 157},  36 (Jan. 2019).

\bibitem{pogorelyuk_laser-guided_2022}
Pogorelyuk, L., Serra, P., Kacker, S., Vlahakis, S., Belsten, N., Rau, G.,
  Carpenter, K.~G., Pueyo, L., Monnier, J.~D., Douglas, E.~S., and Cahoy,
  K.~L., ``Laser-guided space interferometer,'' in [{\em Optical and {{Infrared
  Interferometry}} and {{Imaging VIII}}}{\nolinebreak\hspace{0.1em}]},   {\bf
  12183},  508--521, {SPIE} (Aug. 2022).

\bibitem{Gersh-Range-JWST_thermal}
Gersh-Range, J. and Perrin, M.~D., ``Improving active space telescope wavefront
  control using predictive thermal modeling,'' {\em J. Astron. Telesc. Instrum.
  Syst.}~{\bf 1}(1),  014004 (2015).

\bibitem{mendillo_flight_2012}
Mendillo, C.~B., Chakrabarti, S., Cook, T.~A., Hicks, B.~A., and Lane, B.~F.,
  ``Flight demonstration of a milliarcsecond pointing system for direct
  exoplanet imaging,'' {\em Appl. Opt.}~{\bf 51},  7069--7079 (Oct. 2012).

\bibitem{douglas_wavefront_2018}
Douglas, E.~S., Mendillo, C.~B., Cook, T.~A., Cahoy, K.~L., and Chakrabarti,
  S., ``Wavefront sensing in space: flight demonstration {{II}} of the
  {{PICTURE}} sounding rocket payload,'' {\em Journal of Astronomical
  Telescopes, Instruments, and Systems}~{\bf 4},  019003 (Jan. 2018).

\bibitem{cahoy_wavefront_2013-1}
Cahoy, K.~L., Marinan, A.~D., Novak, B., Kerr, C., Nguyen, T., Webber, M.,
  Falkenburg, G., and Barg, A., ``Wavefront control in space with {{MEMS}}
  deformable mirrors for exoplanet direct imaging,'' {\em J. Micro/Nanolith.
  MEMS MOEMS}~{\bf 13}(1),  011105--011105 (2013).

\bibitem{douglas_small_2021}
Douglas, E., Allan, G., Morgan, R., Holden, B.~G., Gubner, J., Haughwout, C.,
  {do Vale Pereira}, P., Xin, Y., Merk, J., and Cahoy, K.~L., ``Small
  {{Mirrors}} for {{Small Satellites}}: {{Design}} of the {{Deformable Mirror
  Demonstration Mission CubeSat}} ({{DeMi}}) {{Payload}},'' {\em Front. Astron.
  Space Sci.}~{\bf 0} (2021).

\bibitem{morgan_-orbit_2022}
Morgan, R.~E., Vlahakis, S., Douglas, E., Allan, G., Pereira, P. d.~V., Egan,
  M., Furesz, G., Gubner, J., Haughwout, C., Holden, B., Merk, J., Murphy, T.,
  Pogorelyuk, L., Roascio, D., Xin, Y., and Cahoy, K., ``On-orbit operations
  summary for the {{Deformable Mirror Demonstration Mission}} ({{DeMi}})
  {{CubeSat}},'' in [{\em Adaptive {{Optics Systems
  VIII}}}{\nolinebreak\hspace{0.1em}]},   {\bf 12185},  2423--2432, {SPIE}
  (Aug. 2022).

\bibitem{perrin_simulating_2012}
Perrin, M.~D., Soummer, R., Elliott, E.~M., Lallo, M.~D., and Sivaramakrishnan,
  A., ``Simulating point spread functions for the {{James Webb Space
  Telescope}} with {{WebbPSF}},'' in [{\em Proc.
  {{SPIE}}}{\nolinebreak\hspace{0.1em}]},   {\bf 8442},  84423D--84423D--11
  (2012).

\bibitem{greenbaum_-focus_2016}
Greenbaum, A.~Z. and Sivaramakrishnan, A., ``In-focus wavefront sensing using
  non-redundant mask-induced pupil diversity,'' {\em Opt. Express, OE}~{\bf
  24},  15506--15521 (July 2016).

\bibitem{mennesson_roman_2022}
Mennesson, B., Bailey, V.~P., Zellem, R., Hildebrandt, S., Ygouf, M., Rhodes,
  J., Zimmerman, N., Nemati, B., Gonzalez, G., Cady, E., Kern, B., Koch, T.,
  Krist, J., Heydorff, K., Luchik, T., Mok, F., Morrissey, P., Poberezhskiy,
  I., Riggs, A.~J., Shi, F., Zhao, F., Akeson, R., Armus, L., Greenbaum, A.,
  Ingalls, J., and Lowrance, P., ``The {{Roman Space Telescope}} coronagraph
  technology demonstration: current status and relevance to future missions,''
  in [{\em Space {{Telescopes}} and {{Instrumentation}} 2022: {{Optical}},
  {{Infrared}}, and {{Millimeter Wave}}}{\nolinebreak\hspace{0.1em}]},   {\bf
  12180},  685--697, {SPIE} (Aug. 2022).

\bibitem{poberezhskiy_roman_2021}
Poberezhskiy, I., Luchik, T., Zhao, F., Frerking, M., Basinger, S., Cady, E.,
  Colavita, M.~M., Creager, B., Fathpour, N., Goullioud, R., Groff, T.,
  Morrissey, P., Kempenaar, J., Kern, B., Koch, T., Krist, J., Mok, F.,
  Muliere, D., Nemati, B., Riggs, A.~J., Seo, B.-J., Shi, F., Shreckengost, B.,
  Steeves, J., and Tang, H., ``Roman space telescope coronagraph: engineering
  design and operating concept,'' in [{\em Space {{Telescopes}} and
  {{Instrumentation}} 2020: {{Optical}}, {{Infrared}}, and {{Millimeter
  Wave}}}{\nolinebreak\hspace{0.1em}]},   {\bf 11443},  114431V, {International
  Society for Optics and Photonics} (Jan. 2021).

\bibitem{CGISim_manual}
Krist, J., ``{CGISim (Roman coronagraph) Simulator: Public version (uses
  synthetic instead of measured primary \& secondary optic error maps),
  V3.1}.'' \url{https://sourceforge.net/projects/cgisim/} (2022).

\bibitem{ref:GershRange_SPC}
Gersh-Range, J., Riggs, A.~J., and Kasdin, N.~J., ``{Flight designs and pupil
  error mitigation for the bowtie shaped pupil coronagraph on the Nancy Grace
  Roman Space Telescope},'' {\em J. Astron. Telesc. Instrum. Syst.}~{\bf 8}(2),
   025003 (2022).

\bibitem{ref:GershRange_OS11tool}
Gersh-Range, J., ``{Publicly available tools and the overall process for
  generating Observing Scenario (OS) 11 polarization datasets for the
  wide-field-of-view shaped pupil coronagraph}.''
  \url{https://roman.ipac.caltech.edu/docs/Tools_and_Process.pdf} (2022).

\bibitem{dean_phase_2006}
Dean, B.~H., Aronstein, D.~L., Smith, J.~S., Shiri, R., and Acton, D.~S.,
  ``Phase retrieval algorithm for {{JWST Flight}} and {{Testbed Telescope}},''
  in [{\em Space {{Telescopes}} and {{Instrumentation I}}: {{Optical}},
  {{Infrared}}, and {{Millimeter}}}{\nolinebreak\hspace{0.1em}]},   {\bf 6265},
   314--330, {SPIE} (June 2006).

\bibitem{perrin_james_2014}
Perrin, M.~D., Soummer, R., Choquet, {\'E}., N'Diaye, M., Levecq, O., Lajoie,
  C.-P., Ygouf, M., Leboulleux, L., Egron, S., Anderson, R., Long, C., Elliott,
  E., Hartig, G., Pueyo, L., {van der Marel}, R., and Mountain, M., ``James
  {{Webb Space Telescope Optical Simulation Testbed I}}: {{Overview}} and
  {{First Results}},'' {\em arXiv:1407.0591 [astro-ph]} ,  914309 (Aug. 2014).

\bibitem{lajoie_year_2023}
Lajoie, C.-P., Lallo, M., Mel{\'e}ndez, M., Flagey, N., Telfer, R., Comeau,
  T.~M., Kulp, B.~A., Beck, T., Brady, G.~R., and Perrin, M.~D., ``A {{Year}}
  of {{Wavefront Sensing}} with {{JWST}} in {{Flight}}: {{Cycle}} 1 {{Telescope
  Monitoring}} and {{Maintenance Summary}},'' (July 2023).

\bibitem{williams_boost-phase_2022}
Williams, I., Dahlgren, M., Roberts, T.~G., and Karako, T., ``Boost-{{Phase
  Missile Defense}},'' tech. rep., {Center for Strategic and International
  Studies}, {Washington, D.C.} (Fri, 06/24/2022 - 12:00).

\bibitem{scoles_prime_2022}
Scoles, S., ``Prime mover,'' {\em Science}~{\bf 377},  702--705 (Aug. 2022).

\bibitem{chang_spacex_2023}
Chang, K., ``{{SpaceX Starship Test Launch}}: {{Highlights From SpaceX}}'s
  {{Scrubbed Starship Rocket Launch Attempt}},'' {\em The New York Times}
  (Apr. 2023).

\bibitem{fossum_invention_2020}
Fossum, E.~R., ``The {{Invention}} of {{CMOS Image Sensors}}: {{A Camera}} in
  {{Every Pocket}},'' in [{\em 2020 {{Pan Pacific Microelectronics Symposium}}
  ({{Pan Pacific}})}{\nolinebreak\hspace{0.1em}]},   1--6 (Feb. 2020).

\bibitem{alarcon_scientific_2023}
Alarcon, M.~R., Licandro, J., {Serra-Ricart}, M., Joven, E., Gaitan, V., and
  de~Sousa, R., ``Scientific {{CMOS Sensors}} in {{Astronomy}}: {{IMX455}} and
  {{IMX411}},'' {\em PASP}~{\bf 135},  055001 (May 2023).

\bibitem{Goiffon_Vincent}
Goiffon, V.,  [{\em Radiation Effects on CMOS Active Pixel Image
  Sensors}{\nolinebreak\hspace{0.1em}]},  295–332 (11 2015).

\bibitem{leppinen_current_2017}
Leppinen, H., ``Current use of linux in spacecraft flight software,'' {\em IEEE
  Aerospace and Electronic Systems Magazine}~{\bf 32},  4--13 (Oct. 2017).

\bibitem{canham_mars_2022}
Canham, T., ``The {{Mars Ingenuity Helicopter}} - {{A Victory}} for
  {{Open-Source Software}},'' in [{\em 2022 {{IEEE Aerospace Conference}}
  ({{AERO}})}{\nolinebreak\hspace{0.1em}]},   01--11 (Mar. 2022).

\bibitem{knapp_demonstrating_2020}
Knapp, M., Seager, S., Demory, B.-O., Krishnamurthy, A., Smith, M.~W., Pong,
  C.~M., Bailey, V.~P., Donner, A., Pasquale, P.~D., Campuzano, B., Smith, C.,
  Luu, J., Babuscia, A., Robert L.~Bocchino, {\relax Jr}., Loveland, J.,
  Colley, C., Gedenk, T., Kulkarni, T., Hughes, K., White, M., Krajewski, J.,
  and Fesq, L., ``Demonstrating {{High-precision Photometry}} with a
  {{CubeSat}}: {{ASTERIA Observations}} of 55 {{Cancri}} e,'' {\em AJ}~{\bf
  160},  23 (June 2020).

\bibitem{smith_-orbit_2018}
Smith, M., Donner, A., Knapp, M., Pong, C., Smith, C., Luu, J., Pasquale,
  P.~D., and Campuzano, B., ``On-{{Orbit Results}} and {{Lessons Learned}} from
  the {{ASTERIA Space Telescope Mission}},''  20 (2018).

\bibitem{ricker_transiting_2014}
Ricker, G.~R., Winn, J.~N., Vanderspek, R., Latham, D.~W., Bakos, G.~{\'A}.,
  Bean, J.~L., {Berta-Thompson}, Z.~K., Brown, T.~M., Buchhave, L., Butler,
  N.~R., Butler, R.~P., Chaplin, W.~J., Charbonneau, D.~B.,
  {Christensen-Dalsgaard}, J., Clampin, M., Deming, D., Doty, J.~P., Lee,
  N.~D., Dressing, C., Dunham, E.~W., Endl, M., Fressin, F., Ge, J., Henning,
  T., Holman, M.~J., Howard, A.~W., Ida, S., Jenkins, J.~M., Jernigan, G.,
  Johnson, J.~A., Kaltenegger, L., Kawai, N., Kjeldsen, H., Laughlin, G.,
  Levine, A.~M., Lin, D., Lissauer, J.~J., MacQueen, P., Marcy, G., McCullough,
  P.~R., Morton, T.~D., Narita, N., Paegert, M., Palle, E., Pepe, F., Pepper,
  J., Quirrenbach, A., Rinehart, S.~A., Sasselov, D., Sato, B., Seager, S.,
  Sozzetti, A., Stassun, K.~G., Sullivan, P., Szentgyorgyi, A., Torres, G.,
  Udry, S., and Villasenor, J., ``Transiting {{Exoplanet Survey Satellite}},''
  {\em JATIS, JATIAG}~{\bf 1},  014003 (Oct. 2014).

\bibitem{parker_transiting_2018}
Parker, J. J.~K., Lebois, R.~L., Lutz, S., Nickel, C., Ferrant, K., and
  Michaels, A., ``Transiting {{Exoplanet Survey Satellite}} ({{TESS}}) flight
  dynamics commissioning results and experiences,'' {\em AAS GUIDANCE \&
  CONTROL CONFERENCE}  (2018).

\bibitem{gangestad_high_2013}
Gangestad, J.~W., Henning, G.~A., Persinger, R.~R., and Ricker, G.~R., ``A
  {{High Earth}}, {{Lunar Resonant Orbit}} for {{Lower Cost Space Science
  Missions}},'' (Aug. 2013).

\bibitem{oswalt_honeycomb_2013}
Hill, J., Martin, H., and Angel, R., ``Honeycomb {{Mirrors}} for {{Large
  Telescopes}},'' in [{\em Planets, {{Stars}} and {{Stellar
  Systems}}}{\nolinebreak\hspace{0.1em}]},  Oswalt, T.~D. and McLean, I.~S.,
  eds.,  137--184, {Springer Netherlands}, {Dordrecht} (2013).

\bibitem{schlegel_astro2020_2019}
Schlegel, D.~J., Kollmeier, J.~A., Aldering, G., Bailey, S., Baltay, C., Bebek,
  C., BenZvi, S., Besuner, R., Blanc, G., Bolton, A.~S., Bouri, M., Brooks, D.,
  {Buckley-Geer}, E., Cai, Z., Crane, J., Dey, A., Doel, P., Fan, X., Ferraro,
  S., {Font-Ribera}, A., Gutierrez, G., Guy, J., Heetderks, H., Huterer, D.,
  Infante, L., Jelinsky, P., Johns, M., Karagiannis, D., Kent, S.~M., Kim,
  A.~G., Kneib, J.-P., Kronig, L., Konidaris, N., Lahav, O., Lampton, M.~L.,
  Lang, D., Leauthaud, A., Liguori, M., Linder, E.~V., Magneville, C., Martini,
  P., Mateo, M., McDonald, P., Miller, C.~J., Moustakas, J., Myers, A.~D.,
  Mulchaey, J., Newman, J.~A., Nugent, P.~E., {Palanque-Delabrouille}, N.,
  Padmanabhan, N., Piro, A.~L., Poppett, C., Prochaska, J.~X., Pullen, A.~R.,
  Rabinowitz, D., Ramirez, S., Rix, H.-W., Ross, A.~J., Samushia, L., Schaan,
  E., Schubnell, M., Seljak, U., Seo, H.-J., Shectman, S.~A., Silber, J.,
  Simon, J.~D., Slepian, Z., {Soares-Santos}, M., Tarle, G., Thompson, I.,
  Valluri, M., Wechsler, R.~H., White, M., Wilson, M.~J., Yeche, C., and
  Zaritsky, D., ``Astro2020 {{APC White Paper}}: {{The MegaMapper}}: a z {$>$}
  2 spectroscopic instrument for the study of {{Inflation}} and {{Dark
  Energy}},'' (July 2019).

\bibitem{korsch_anastigmatic_1977}
Korsch, D., ``Anastigmatic three-mirror telescope,'' {\em Appl. Opt., AO}~{\bf
  16},  2074--2077 (Aug. 1977).

\bibitem{feinberg_space_2012}
Feinberg, L.~D., Dean, B.~H., Hayden, W.~L., Howard, J.~M., {Keski-Kuha},
  R.~A., and Cohen, L.~M., ``Space telescope design considerations,'' {\em
  OE}~{\bf 51},  011006 (Feb. 2012).

\bibitem{roddier_combined_1993}
Roddier, C. and Roddier, F., ``Combined approach to the {{Hubble Space
  Telescope}} wave-front distortion analysis,'' {\em Applied optics}~{\bf
  32}(16),  2992--3008 (1993).

\bibitem{litvak_image_1991}
Litvak, M.~M., ``Image inversion analysis of the {{HST OTA}} ({{Hubble Space
  Telescope Optical Telescope Assembly}}), phase {{A}},'' Tech. Rep.
  JPL-9950-1381 (July 1991).

\bibitem{krist_phase-retrieval_1995}
Krist, J.~E. and Burrows, C.~J., ``Phase-retrieval analysis of pre- and
  post-repair {{Hubble Space Telescope}} images,'' {\em Applied Optics}~{\bf
  34},  4951 (Aug. 1995).

\bibitem{schlawin_jwst_2023}
Schlawin, E., Beatty, T., Brooks, B., Nikolov, N.~K., Greene, T.~P., Espinoza,
  N., Glidic, K., Baka, K., Egami, E., Stansberry, J., Boyer, M., Gennaro, M.,
  Leisenring, J., Hilbert, B., Misselt, K., Kelly, D., Canipe, A., Beichman,
  C., Correnti, M., Knight, J.~S., Jurling, A., Perrin, M.~D., Feinberg, L.~D.,
  McElwain, M.~W., Bond, N., Ciardi, D., Kendrew, S., and Rieke, M., ``{{JWST
  NIRCam Defocused Imaging}}: {{Photometric Stability Performance}} and {{How
  It Can Sense Mirror Tilts}},'' {\em PASP}~{\bf 135},  018001 (Jan. 2023).

\bibitem{nguyen_fine-pointing_2018}
Nguyen, T., Morgan, E., Vanderspek, R., Levine, A., Kephart, M., Francis, J.,
  Zapetis, J., Cahoy, K., and Jr, G.~R., ``Fine-pointing performance and
  corresponding photometric precision of the {{Transiting Exoplanet Survey
  Satellite}},'' {\em JATIS}~{\bf 4},  047001 (Sept. 2018).

\bibitem{bartusek_nancy_2022}
Bartusek, L.~M., Davis, J.~L., and Vess, M.~F., ``Nancy {{Grace Roman Space
  Telescope Observatory Implementation}} and {{Challenges}},'' in [{\em 2022
  {{IEEE Aerospace Conference}} ({{AERO}})}{\nolinebreak\hspace{0.1em}]},
  01--14 (Mar. 2022).

\bibitem{parnin_top_2017}
Parnin, C., Helms, E., Atlee, C., Boughton, H., Ghattas, M., Glover, A.,
  Holman, J., Micco, J., Murphy, B., Savor, T., Stumm, M., Whitaker, S., and
  Williams, L., ``The {{Top}} 10 {{Adages}} in {{Continuous Deployment}},''
  {\em IEEE Software}~{\bf 34},  86--95 (May 2017).

\bibitem{tibazarwa_strategic_2021}
Tibazarwa, A., ``Strategic {{Integration}} for {{Hardware}} and {{Software
  Convergence Complexity}},'' {\em IEEE Engineering Management Review}~{\bf
  49}(3),  92--102 (2021).

\bibitem{omullane22a}
O'Mullane, W., Economou, F., Lim, K.-T., Mueller, F., Jenness, T.,
  Dubois-Felsmann, G.~P., Guy, L.~P., Sullivan, I.~S., AlSayyad, Y., Swinbank,
  J.~D., and Krughoff, K.~S., ``Software architecture and system design of
  rubin observatory,'' (2022).

\bibitem{holm_making_2002}
Holm, J., ``Making {{Sense}} of {{Rocket Science}} - {{Building NASA}}'s
  {{Knowledge Management Program}},'' in [{\em Hawaii {{International
  Conference}} on {{System Sciences}}}{\nolinebreak\hspace{0.1em}]},  (Jan.
  2002).

\bibitem{douglas_wfirst_2018}
Douglas, E.~S., Carlton, A.~K., Cahoy, K.~L., Kasdin, N.~J., Turnbull, M., and
  Macintosh, B., ``{{WFIRST}} coronagraph technology requirements: status
  update and systems engineering approach,'' in [{\em Modeling, {{Systems
  Engineering}}, and {{Project Management}} for {{Astronomy
  VIII}}}{\nolinebreak\hspace{0.1em}]},   {\bf 10705},  1070526, {International
  Society for Optics and Photonics} (July 2018).

\bibitem{browning_doorstop:_2014}
Browning, J. and Adams, R., ``Doorstop: {{Text-Based Requirements Management
  Using Version Control}},'' {\em Journal of Software Engineering and
  Applications}~{\bf 07}(03),  187--194 (2014).

\bibitem{olbert_casting_1994}
Olbert, B.~H., Angel, J. R.~P., Hill, J.~M., and Hinman, S.~F., ``Casting
  6.5-meter mirrors for the {{MMT}} conversion and {{Magellan}},'' in [{\em
  Advanced {{Technology Optical Telescopes V}}}{\nolinebreak\hspace{0.1em}]},
  {\bf 2199},  144--155, {SPIE} (June 1994).

\bibitem{martin_fabrication_1997}
Martin, H.~M., Burge, J.~H., Ketelsen, D.~A., and West, S.~C., ``Fabrication of
  the 6.5-m primary mirror for the {{Multiple Mirror Telescope Conversion}},''
  in [{\em Optical {{Telescopes}} of {{Today}} and
  {{Tomorrow}}}{\nolinebreak\hspace{0.1em}]},  Ardeberg, A.~L., ed.,  399--404
  (Mar. 1997).

\bibitem{geyl_polishing_1999}
Martin, H.~M., Allen, R.~G., Burge, J.~H., Dettmann, L.~R., Ketelsen, D.~A.,
  Kittrell, W.~C., and Miller, S.~M., ``Polishing of a 6.5-m f/1.25 mirror for
  the first {{Magellan}} telescope,'' in [{\em Optical {{Systems Design}} and
  {{Production}}}{\nolinebreak\hspace{0.1em}]},  Geyl, R. and Maxwell, J.,
  eds.,  47 (Sept. 1999).

\bibitem{kingsley_inexpensive_2018}
Kingsley, J.~S., Angel, R., Davison, W., Neff, D., Teran, J., Assenmacher, B.,
  Peyton, K., Martin, H.~M., Oh, C., Kim, D., Pearce, E., Rascon, M., Connors,
  T., Alfred, D., Jannuzi, B.~T., Zaritsky, D., Christensen, E., Males, J.,
  Hinz, P., Seaman, R., Gonzales, K., and Adriaanse, D., ``An inexpensive
  turnkey 6.5m observatory with customizing options,'' in [{\em Ground-based
  and {{Airborne Telescopes VII}}}{\nolinebreak\hspace{0.1em}]},   {\bf 10700},
   107004H, {International Society for Optics and Photonics} (July 2018).

\bibitem{miyata_university_2022}
Miyata, T., Yoshii, Y., Doi, M., Kohno, K., Tanaka, M., Motohara, K., Minezaki,
  T., Sako, S., Morokuma, T., Tanabe, T., Hatsukade, B., Konishi, M.,
  Takahashi, H., Kamizuka, T., Egusa, F., Sameshima, H., Asano, K., Nishimura,
  A., Koyama, S., Kato, N., Numata, M., Aoki, T., Bronfman, L., Ruiz, M.,
  Hamuy, M., Mendez, R., Garay, G., and Escala, A., ``The {{University}} of
  {{Tokyo Atacama Observatory}} 6.5m telescope: project status 2022,'' in [{\em
  Ground-based and {{Airborne Telescopes IX}}}{\nolinebreak\hspace{0.1em}]},
  {\bf 12182},  385--393, {SPIE} (Aug. 2022).

\bibitem{kim_compact_2023}
Kim, D. and {et al}, ``Compact {{Three Mirror Anastigmat Space Telescope
  Design}} using 6.5m {{Monolithic Primary Mirror}},'' in [{\em Proc
  {{SPIE}}}{\nolinebreak\hspace{0.1em}]},   {\bf 12677} (2023).

\bibitem{choi_approaches_2023}
Choi, H. and {et al}, ``Approaches to developing tolerance and error budget for
  active three mirror anastigmat space telescopes,'' in [{\em Proc
  {{SPIE}}}{\nolinebreak\hspace{0.1em}]},   {\bf 12677} (2023).

\bibitem{derby_integrated_2023}
Derby, K.~Z. and {et al}, ``Integrated modeling of wavefront sensing and
  control for space telescopes utilizing active and adaptive optics,'' in [{\em
  Proc {{SPIE}}}{\nolinebreak\hspace{0.1em}]},   {\bf 12677} (2023).

\bibitem{kang_focus_2023}
Kang, H. and {et al}, ``Focus diverse phase retrieval testbed development of
  continuous wavefront sensing for space telescope applications,'' in [{\em
  Proc {{SPIE}}}{\nolinebreak\hspace{0.1em}]},   {\bf 12677} (2023).

\bibitem{Blomquist_analysis_2023}
Blomquist, S. and {et al}, ``Analysis of {{Active Optics Correction}} for a
  6.5m {{Honeycomb Mirror}} in a {{Space Observatory}},'' in [{\em Proc
  {{SPIE}}}{\nolinebreak\hspace{0.1em}]},   {\bf 12677} (2023).

\bibitem{nurre_hubble_1989}
Nurre, G.~S., Anhouse, S.~J., and Gullapalli, S.~N., ``Hubble {{Space Telescope
  Fine Guidance Sensor Control System}},'' in [{\em Acquisition, {{Tracking}},
  and {{Pointing III}}}{\nolinebreak\hspace{0.1em}]},   {\bf 1111},  327--343,
  {International Society for Optics and Photonics} (Sept. 1989).

\bibitem{bahcall_universe_1980}
Bahcall, J.~N. and Soneira, R.~M., ``The universe at faint magnitudes.
  {{I-Models}} for the galaxy and the predicted star counts,'' {\em The
  Astrophysical Journal Supplement Series}~{\bf 44},  73--110 (1980).

\bibitem{elliott_5_2021}
Elliott, A., Richardson, N.~D., Pablo, H., Moffat, A. F.~J., Bowman, D.~M.,
  Ibrahim, N., Handler, G., Lovekin, C., Popowicz, A., {St-Louis}, N., Wade,
  G.~A., and Zwintz, K., ``5 yr of {{BRITE-Constellation}} photometry of the
  luminous blue variable {{P Cygni}}: properties of the stochastic
  low-frequency variability,'' {\em Monthly Notices of the Royal Astronomical
  Society}~{\bf 509},  4246--4255 (Jan. 2021).

\bibitem{perlmutter_key_2019}
Perlmutter, S., Aldering, G., Baltay, C., Deustua, S., Freedman, W., Fruchter,
  A., Rubin, D., Sako, M., and Suntzeff, N., ``The {{Key Role}} of {{Supernova
  Spectrophotometry}} in the {{Next-Decade Dark Energy Science Program}},''
  {\em Bulletin of the American Astronomical Society}~{\bf 51},  494 (May
  2019).

\bibitem{boone_2021a}
{Boone}, K., {Aldering}, G., {Antilogus}, P., {Aragon}, C., {Bailey}, S.,
  {Baltay}, C., {Bongard}, S., {Buton}, C., {Copin}, Y., {Dixon}, S.,
  {Fouchez}, D., {Gangler}, E., {Gupta}, R., {Hayden}, B., {Hillebrandt}, W.,
  {Kim}, A.~G., {Kowalski}, M., {K{\"u}sters}, D., {L{\'e}get}, P.~F.,
  {Mondon}, F., {Nordin}, J., {Pain}, R., {Pecontal}, E., {Pereira}, R.,
  {Perlmutter}, S., {Ponder}, K.~A., {Rabinowitz}, D., {Rigault}, M., {Rubin},
  D., {Runge}, K., {Saunders}, C., {Smadja}, G., {Suzuki}, N., {Tao}, C.,
  {Taubenberger}, S., {Thomas}, R.~C., and {Vincenzi}, M., ``{The Twins
  Embedding of Type Ia Supernovae. I. The Diversity of Spectra at Maximum
  Light},'' {\em ApJ}~{\bf 912},  70 (May 2021).

\bibitem{boone_2021b}
{Boone}, K., {Aldering}, G., {Antilogus}, P., {Aragon}, C., {Bailey}, S.,
  {Baltay}, C., {Bongard}, S., {Buton}, C., {Copin}, Y., {Dixon}, S.,
  {Fouchez}, D., {Gangler}, E., {Gupta}, R., {Hayden}, B., {Hillebrandt}, W.,
  {Kim}, A.~G., {Kowalski}, M., {K{\"u}sters}, D., {L{\'e}get}, P.~F.,
  {Mondon}, F., {Nordin}, J., {Pain}, R., {Pecontal}, E., {Pereira}, R.,
  {Perlmutter}, S., {Ponder}, K.~A., {Rabinowitz}, D., {Rigault}, M., {Rubin},
  D., {Runge}, K., {Saunders}, C., {Smadja}, G., {Suzuki}, N., {Tao}, C.,
  {Taubenberger}, S., {Thomas}, R.~C., and {Vincenzi}, M., ``{The Twins
  Embedding of Type Ia Supernovae. II. Improving Cosmological Distance
  Estimates},'' {\em ApJ}~{\bf 912},  71 (May 2021).

\bibitem{peters-limbach_optical_2013}
{Peters-Limbach}, M.~A., Groff, T.~D., Kasdin, N.~J., Driscoll, D., Galvin, M.,
  Foster, A., Carr, M.~A., LeClerc, D., Fagan, R., McElwain, M.~W., Knapp, G.,
  Brandt, T., Janson, M., Guyon, O., Jovanovic, N., Martinache, F., Hayashi,
  M., and Takato, N., ``The optical design of {{CHARIS}}: an exoplanet {{IFS}}
  for the {{Subaru}} telescope,'' in [{\em Proc.
  {{SPIE}}}{\nolinebreak\hspace{0.1em}]},   {\bf 8864},  88641N--88641N--15
  (2013).

\bibitem{claudi_sphere_2010}
Claudi, R.~U., Turatto, M., Giro, E., Mesa, D., Anselmi, U., Bruno, P.,
  Cascone, E., De~Caprio, V., Desidera, S., Dorn, R., Fantinel, D., Finger, G.,
  Gratton, R.~G., Lessio, L., Lizon, J.~L., Salasnic, B., Scuderi, S., Dohlen,
  {\relax Kj}., Beuzit, J.~L., Puget, P., Antichi, J., Hubin, N., and Kasper,
  M., ``{{SPHERE IFS}}: the spectro differential imager of the {{VLT}} for
  exoplanets search,''  {\bf 7735},  77350V--77350V--11 (2010).

\bibitem{lantz_2004}
{Lantz}, B., {Aldering}, G., {Antilogus}, P., {Bonnaud}, C., {Capoani}, L.,
  {Castera}, A., {Copin}, Y., {Dubet}, D., {Gangler}, E., {Henault}, F.,
  {Lemonnier}, J.-P., {Pain}, R., {Pecontal}, A., {Pecontal}, E., and {Smadja},
  G., ``{SNIFS: a wideband integral field spectrograph with microlens
  arrays},'' in [{\em Optical Design and
  Engineering}{\nolinebreak\hspace{0.1em}]},  {Mazuray}, L., {Rogers}, P.~J.,
  and {Wartmann}, R., eds., {\em Society of Photo-Optical Instrumentation
  Engineers (SPIE) Conference Series} {\bf 5249},  146--155 (Feb. 2004).

\bibitem{bacon_2010}
{Bacon}, e.~a., ``{The MUSE second-generation VLT instrument},'' in [{\em
  Ground-based and Airborne Instrumentation for Astronomy
  III}{\nolinebreak\hspace{0.1em}]},  {McLean}, I.~S., {Ramsay}, S.~K., and
  {Takami}, H., eds., {\em Society of Photo-Optical Instrumentation Engineers
  (SPIE) Conference Series} {\bf 7735},  773508 (July 2010).

\bibitem{Prieto_2008}
{Prieto}, E., {Ealet}, A., {Milliard}, B., {Aumeunier}, M.-H., {Bonissent}, A.,
  {Cerna}, C., {Crouzet}, P.-E., {Karst}, P., {Kneib}, J.-P., {Malina}, R.,
  {Pamplona}, T., {Rossin}, C., {Smadja}, G., and {Vives}, S., ``{An integral
  field spectrograph for SNAP},'' in [{\em Space Telescopes and Instrumentation
  2008: Optical, Infrared, and Millimeter}{\nolinebreak\hspace{0.1em}]},
  {Oschmann}, Jacobus~M., J., {de Graauw}, M. W.~M., and {MacEwen}, H.~A.,
  eds., {\em Society of Photo-Optical Instrumentation Engineers (SPIE)
  Conference Series} {\bf 7010},  701019 (July 2008).

\bibitem{WFIRST_IFS}
{Gao}, G., {Pasquale}, B.~A., {Marx}, C.~T., and {Chambers}, V.~J., ``{Optical
  design of the WFIRST Phase-A Integral Field Channel},'' in [{\em Society of
  Photo-Optical Instrumentation Engineers (SPIE) Conference
  Series}{\nolinebreak\hspace{0.1em}]},  {Clark}, P.~P., {Muschaweck}, J.~A.,
  {Pfisterer}, R.~N., and {Rogers}, J.~R., eds., {\em Society of Photo-Optical
  Instrumentation Engineers (SPIE) Conference Series} {\bf 10590},  105901R
  (Nov. 2017).

\bibitem{mendillo_picture-c_2019}
Mendillo, C.~B., Hewawasam, K., Howe, G.~A., Martel, J., Cook, T.~A., and
  Chakrabarti, S., ``The {{PICTURE-C}} exoplanetary direct imaging balloon
  mission: first flight preparation,'' in [{\em Techniques and
  {{Instrumentation}} for {{Detection}} of {{Exoplanets
  IX}}}{\nolinebreak\hspace{0.1em}]},   {\bf 11117},  101--111, {SPIE} (Sept.
  2019).

\bibitem{maier_design_2020}
Maier, E.~R., Douglas, E.~S., Kim, D.~W., Su, K., Ashcraft, J.~N.,
  Breckinridge, J.~B., Choi, H., Choquet, E., Connors, T.~E., Durney, O.,
  Gonzales, K.~L., Guthery, C.~E., Haughwout, C.~A., Heath, J.~C., Hyatt, J.,
  Lumbres, J., Males, J.~R., Matthews, E.~C., Milani, K., Montoya, O.~M.,
  N'Diaye, M., Noenickx, J., Pogorelyuk, L., Ruane, G., Schneider, G., Smith,
  G.~A., and Stark, C.~C., ``Design of the vacuum high contrast imaging testbed
  for {{CDEEP}}, the {{Coronagraphic Debris}} and {{Exoplanet Exploring
  Pioneer}},'' in [{\em Space {{Telescopes}} and {{Instrumentation}} 2020:
  {{Optical}}, {{Infrared}}, and {{Millimeter
  Wave}}}{\nolinebreak\hspace{0.1em}]},   {\bf 11443},  114431Y, {International
  Society for Optics and Photonics} (Dec. 2020).

\bibitem{ashcraft_space_2022}
Ashcraft, J.~N., Choi, H., Douglas, E.~S., Derby, K., Van~Gorkom, K., Kim, D.,
  Anche, R., Carter, A., Durney, O., Haffert, S., Harrison, L., Kautz, M.,
  Lumbres, J., Males, J.~R., Milani, K., Montoya, O.~M., and Smith, G.~A.,
  ``The {{Space Coronagraph Optical Bench}} ({{SCoOB}}): 1. {{Design}} and
  {{Assembly}} of a {{Vacuum-compatible Coronagraph Testbed}} for {{Spaceborne
  High-Contrast Imaging Technology}},'' (Aug. 2022).

\bibitem{van_gorkom_space_2022}
Van~Gorkom, K., Douglas, E.~S., Ashcraft, J.~N., Haffert, S., Kim, D., Choi,
  H., Anche, R.~M., Males, J.~R., Milani, K., Derby, K., Harrison, L., and
  Durney, O., ``The space coronagraph optical bench ({{SCoOB}}): 2. wavefront
  sensing and control in a vacuum-compatible coronagraph testbed for spaceborne
  high-contrast imaging technology,'' (Aug. 2022).

\bibitem{blaurock_structural-thermal-optical_2005}
Blaurock, C., McGinnis, M., Kim, K., and Mosier, G.~E.,
  ``Structural-thermal-optical performance ({{STOP}}) sensitivity analysis for
  the {{James Webb Space Telescope}},'' in [{\em Optical {{Modeling}} and
  {{Performance Predictions II}}}{\nolinebreak\hspace{0.1em}]},   {\bf 5867},
  246--256, {SPIE} (Aug. 2005).

\bibitem{saini_impipeline_2017}
Saini, N., Anderson, K., Chang, Z., Gutt, G., and Nemati, B., ``{{IMPipeline}}:
  an integrated {{STOP}} modeling pipeline for the {{WFIRST}} coronagraph
  ({{Conference Presentation}}),'' in [{\em Techniques and {{Instrumentation}}
  for {{Detection}} of {{Exoplanets VIII}}}{\nolinebreak\hspace{0.1em}]},
  {\bf 10400},  1040008, {International Society for Optics and Photonics} (Oct.
  2017).

\bibitem{ashcraft_versatile_2021}
Ashcraft, J.~N., Douglas, E.~S., Kim, D., Smith, G.~A., Cahoy, K., Connors, T.,
  Derby, K.~Z., Gasho, V., Gonzales, K., Guthery, C.~E., Kim, G.~H., Sauve, C.,
  and Serra, P., ``The versatile {{CubeSat Telescope}}: going to large
  apertures in small spacecraft,'' in [{\em {{UV}}/{{Optical}}/{{IR Space
  Telescopes}} and {{Instruments}}: {{Innovative Technologies}} and {{Concepts
  X}}}{\nolinebreak\hspace{0.1em}]},   {\bf 11819},  1181904, {International
  Society for Optics and Photonics} (Aug. 2021).

\bibitem{males_ground-based_2018}
Males, J.~R. and Guyon, O., ``Ground-based adaptive optics coronagraphic
  performance under closed-loop predictive control,'' {\em JATIS, JATIAG}~{\bf
  4},  019001 (Feb. 2018).

\bibitem{lyon_space_2012}
Lyon, R.~G. and Clampin, M., ``Space telescope sensitivity and controls for
  exoplanet imaging,'' {\em OE, OPEGAR}~{\bf 51},  011002 (Feb. 2012).

\bibitem{gersh-range_improving_2014}
{Gersh-Range}, J. and Perrin, M.~D., ``Improving active space telescope
  wavefront control using predictive thermal modeling,'' {\em JATIS}~{\bf 1},
  014004 (Oct. 2014).

\bibitem{brooks_predictive_2017}
Brooks, T., ``Predictive thermal control applied to {{HabEx}},'' in [{\em
  {{UV}}/{{Optical}}/{{IR Space Telescopes}} and {{Instruments}}: {{Innovative
  Technologies}} and {{Concepts VIII}}}{\nolinebreak\hspace{0.1em}]},  MacEwen,
  H.~A. and Breckinridge, J.~B., eds.,  40, {SPIE}, {San Diego, United States}
  (Sept. 2017).

\bibitem{males_magao-x:_2018}
Males, J.~R., Close, L.~M., Miller, K., Schatz, L., Doelman, D., Lumbres, J.,
  Snik, F., Rodack, A., Knight, J., Gorkom, K.~V., Long, J.~D., Hedglen, A.,
  Kautz, M., Jovanovic, N., Morzinski, K., Guyon, O., Douglas, E., Follette,
  K.~B., Lozi, J., Bohlman, C., Durney, O., Gasho, V., Hinz, P., Ireland, M.,
  Jean, M., Keller, C., Kenworthy, M., Mazin, B., Noenickx, J., Alfred, D.,
  Perez, K., Sanchez, A., Sauve, C., Weinberger, A., and Conrad, A.,
  ``{{MagAO-X}}: project status and first laboratory results,'' in [{\em
  Adaptive {{Optics Systems VI}}}{\nolinebreak\hspace{0.1em}]},   {\bf 10703},
  1070309, {International Society for Optics and Photonics} (July 2018).

\bibitem{guyon_compute_2018}
Guyon, O., Sevin, A., Gratadour, D., Bernard, J., Ltaief, H., Sukkari, D.,
  Cetre, S., Skaf, N., Lozi, J., Martinache, F., Clergeon, C., Norris, B.,
  Wong, A., and Males, J., ``The compute and control for adaptive optics
  ({{CACAO}}) real-time control software package,'' in [{\em Adaptive {{Optics
  Systems VI}}}{\nolinebreak\hspace{0.1em}]},   {\bf 10703},  107031E,
  {International Society for Optics and Photonics} (July 2018).

\bibitem{schatz_three-sided_2022}
Schatz, L., Codona, J., Long, J.~D., Males, J.~R., Pullen, W., Lumbres, J.,
  Van~Gorkom, K., Chambouleyron, V., Close, L.~M., Correia, C., Fauvarque, O.,
  Fusco, T., Guyon, O., Hart, M., {Janin-Potiron}, P., Johnson, R., Jovanovic,
  N., Mateen, M., Sauvage, J.-F., and Neichel, B., ``Three-sided pyramid
  wavefront sensor. {{II}}. {{Preliminary}} demonstration on the new {{CACTI}}
  testbed,'' (Oct. 2022).

\bibitem{weckenborg_improving_2020}
Weckenborg, C., Kieckh{\"a}fer, K., Spengler, T.~S., Bernstein, P., and Hahn,
  M., ``Improving {{Resource Utilisation}} in {{Prototype Vehicle
  Production}},'' {\em Impact}~{\bf 2020},  13--18 (July 2020).

\bibitem{swartwout_cubesat_2015}
Swartwout, M., ``{{CubeSat Mission Success}} (or {{Not}}): {{Trends}} and
  {{Recommendations}},'' (June 2015).

\bibitem{clark_sweet_2008}
Clark, S., ``Sweet success at last for {{Falcon}} 1 rocket,'' (Sept. 2008).

\bibitem{yost_small_2018}
Yost, B.~D., Burkhard, C.~D., Mayer, D.~J., Weston, S.~V., and Fishman, J.~L.,
  ``Small {{Spacecraft Systems Virtual Institute}}'s {{Federated Databases}}
  and {{State}} of the {{Art}} of {{Small Spacecraft Report}},'' in [{\em Small
  {{Satellite Conference Proceedings}}}{\nolinebreak\hspace{0.1em}]},
  SSC18--IV--06 (2018).

\bibitem{bocchino_f_2018}
Bocchino, R., Canham, T., Watney, G., Reder, L., and Levison, J., ``F
  {{Prime}}: {{An Open-Source Framework}} for {{Small-Scale Flight Software
  Systems}},'' {\em AIAA/USU Conference on Small Satellites}  (Aug. 2018).

\bibitem{tan_2023_2023}
Tan, F., ``2023 {{NASA Science}}: {{NASA SmallSats Missions}} for {{Science}}
  and {{Technology Demonstration}},'' (2023).

\bibitem{martin_production_2022}
Martin, H.~M., Ceragioli, R., Gasho, V., Jannuzi, B.~T., Kim, D.~W., Kingsley,
  J.~S., Law, K., Loeff, A., Lutz, R.~D., Meyen, S., Oh, C.~J., Tuell, M.~T.,
  Weinberger, S.~N., West, S.~C., Whitsitt, R., and Wortley, R., ``Production
  of 8.4 m primary mirror segments for {{GMT}},'' in [{\em Advances in
  {{Optical}} and {{Mechanical Technologies}} for {{Telescopes}} and
  {{Instrumentation V}}}{\nolinebreak\hspace{0.1em}]},   {\bf 12188},
  177--184, {SPIE} (Sept. 2022).

\bibitem{martin_active_1998}
Martin, H.~M., Callahan, S.~P., Cuerden, B., Davison, W.~B., DeRigne, S.~T.,
  Dettmann, L.~R., Parodi, G., Trebisky, T.~J., West, S.~C., and Williams,
  J.~T., ``Active supports and force optimization for the {{MMT}} primary
  mirror,'' in [{\em Astronomical {{Telescopes}} \&
  {{Instrumentation}}}{\nolinebreak\hspace{0.1em}]},  Stepp, L.~M., ed.,
  412--423 (Aug. 1998).

\bibitem{martin_design_2006}
Martin, H.~M., Angel, J. R.~P., Burge, J.~H., Cuerden, B., Davison, W.~B.,
  Johns, M., Kingsley, J.~S., Kot, L.~B., Lutz, R.~D., Miller, S.~M., Shectman,
  S.~A., Strittmatter, P.~A., and Zhao, C., ``Design and manufacture of 8.4m
  primary mirror segments and supports for the {{GMT}},'' in [{\em {{SPIE
  Astronomical Telescopes}} +
  {{Instrumentation}}}{\nolinebreak\hspace{0.1em}]},  {Atad-Ettedgui}, E.,
  Antebi, J., and Lemke, D., eds.,  62730E (June 2006).

\bibitem{martin_production_2012}
Martin, H.~M., Allen, R.~G., Burge, J.~H., Kim, D.~W., Kingsley, J.~S., Law,
  K., Lutz, R.~D., Strittmatter, P.~A., Su, P., Tuell, M.~T., West, S.~C., and
  Zhou, P., ``Production of 8.4m segments for the {{Giant Magellan
  Telescope}},'' in [{\em Modern {{Technologies}} in {{Space-}} and
  {{Ground-based Telescopes}} and {{Instrumentation
  II}}}{\nolinebreak\hspace{0.1em}]},   {\bf 8450},  801--815, {SPIE} (Sept.
  2012).

\bibitem{martin_manufacture_2020}
Martin, H.~M., Ceragioli, R., Jannuzi, B., Kim, D.~W., Kingsley, J., Law, K.,
  Loeff, A., Lutz, R., McMahon, T., Meyen, S., Oh, C.~J., Tuell, M.,
  Weinberger, S., West, S., and Wortley, R., ``Manufacture of 8.4 m segments
  for the {{GMT}} primary mirror,'' in [{\em Advances in {{Optical}} and
  {{Mechanical Technologies}} for {{Telescopes}} and {{Instrumentation
  IV}}}{\nolinebreak\hspace{0.1em}]},  Geyl, R. and Navarro, R., eds.,  289,
  {SPIE}, {Online Only, United States} (Dec. 2020).

\bibitem{pridgeon_james_2022}
Pridgeon, A., ``James {{B}}. {{Breckinridge}} (1939\textendash 2022),'' {\em
  Bulletin of the AAS}~{\bf 54} (Jan. 2022).

\end{thebibliography}
